\documentclass[seceqn,dvips]{arxstspdf} 
\usepackage{dcolumn}
\usepackage{multirow}
\usepackage{graphicx}
\usepackage{flushend}
\usepackage{stfloats}
\usepackage{rotate2e}

\volume{25}
\issue{4}
\pubyear{2010}
\firstpage{545}
\lastpage{565}
\doi{10.1214/10-STS342}

\makeatletter
\newcommand{\var}{\operatorname{var}}
\newcolumntype{d}[1]{D{.}{.}{#1}}
\setattribute{ead}{sep}{; }
\makeatother

\begin{document}
\begin{frontmatter}

\title{A Statistical Model to Explain the Mendel--Fisher Controversy}
\runtitle{Explaining the Mendel--Fisher Controversy}

\begin{aug}
\author{\fnms{Ana M.} \snm{Pires}\corref{}\ead[label=e1]{apires@math.ist.utl.pt}}
\and
\author{\fnms{Jo\~ao A.} \snm{Branco}\ead[label=e2]{jbranco@math.ist.utl.pt}}
\runauthor{A. M. Pires and J. A. Branco}
\pdfauthor{Ana M. Pires, Joao A. Branco}

\affiliation{Technical University of Lisbon}

\address{Ana M. Pires is Assistant Professor
and
Jo\~ao A. Branco is Associate Professor,
Department of Mathematics and CEMAT, IST, Technical University of Lisbon (TULisbon),
Av. Rovisco Pais, 1049-001 Lisboa, Portugal
\printead{e1,e2}.}

\end{aug}

%
\begin{abstract}
In 1866 Gregor Mendel published a seminal paper containing the
foundations of modern genetics. In 1936 Ronald Fisher published a
statistical analysis of Mendel's data concluding that ``\textit{the
data of most, if not all, of the experiments have been falsified so as
to agree closely with Mendel's expectations}.'' The accusation gave
rise to a controversy which has reached the present time.
There are reasonable grounds to assume that a certain unconscious bias
was systematically introduced in Mendel's experimentation. Based on
this assumption, a probability model that fits Mendel's data and does
not offend Fisher's analysis is given. This reconciliation model may
well be the end of the Mendel--Fisher controversy.
\end{abstract}

%

\end{frontmatter}

\section{Introduction}\label{sec1}

Gregor Mendel
is recognized as a brilliant scientist and the founder of modern
genetics. However, long ago, another eminent scientist, the
statistician and geneticist, Sir Ronald Fisher, questioned Mendel's
integrity claiming that Mendel's data agree better with his theory than
expected under natural fluctuations.
Fisher's conclusion is based on strong statistical arguments and has
been interpreted as an evidence of misconduct. A large number of papers
about this controversy have been produced, culminating with the
publication in 2008 of a book (Frank\-lin et al., \citeyear{Fra2008}) aimed at
ending the polemic and definitely rehabilitating Mendel's image.
However, the authors recognize, ``\textit{the issue of the `too good to~be true'
aspect of Mendel's data found by Fisher still stands}''.

After submitting Mendel's data and Fisher's statistical analysis to
extensive computations and Monte Carlo simulations, attempting to
discover a hidden explanation that could either confirm or refute
Fisher's allegation,
we have concluded that a statistical model with a simple probability
mechanism can clarify the controversy, that is, explain Fisher's
conclusions without accusing Mendel (or any assistant) of deliberate
fraud.

The paper is organized as follows. In Section~\ref{sec2} we summarize the
history of the controversy.
Then, in Section~\ref{sec3}, we present a brief description of Mendel's
experiments and of the data under consideration. In Section~\ref{sec4} we
examine previous statistical analyses of Mendel's data, including
Fisher's chi-square analysis and a meta-analysis of $p$-values. In
Section~\ref{sec5} we present the proposed statistical model and show how it can
explain the pending issues. The conclusions of this work are summed up
in Section~\ref{sec6}.

\section{A Brief History of the Mendel--Fisher Controversy}\label{sec2}

To situate the reader within the context of the subject matter, we
first highlight the most significant characteristics of the two leading
figures and review the key aspects and chronology of the controversy.

\begin{figure}
\centering
\begin{tabular}{cc}

\includegraphics{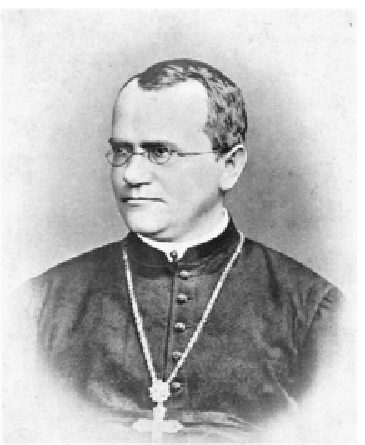}
&\includegraphics{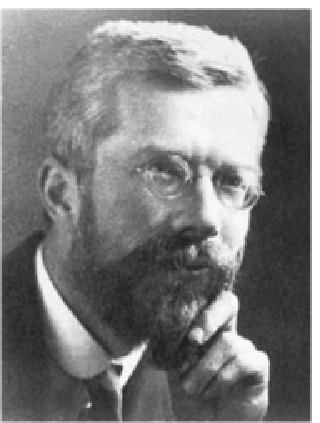}\\
\footnotesize{(a)}&\footnotesize{(b)}
\end{tabular}
\caption{\textup{(a)} Mendel's portrait which appeared as frontispiece in the
book \textit{Mendel's Principles of Heredity, a Defense}, Bateson (\citeyear{Bat1902}).
\textup{(b)} A picture of Sir R. A. Fisher.}
\label{Mendel}
\end{figure}

Gregor Mendel [1822--1884, Figure~\ref{Mendel}(a)] was an Augustinian
Austrian monk who, during at least seven years, performed controlled
crossing experiments with the garden pea (\textit{Pisum sativum} L.).
He may have personally controlled the fertilization of around 29,000 plants.
Based on the results of these experiments, he formulated the two laws,
or principles, of heredity (Mendel's first law: principle of
segregation; Mendel's second law: principle of independent assortment).
Mendel's findings were published in 1866 in the Proceedings of the
Society of Natural History of Br\"unn, Mendel (\citeyear{Me1866}).
To draw his conclusions, Mendel analyzed the data informally, that is,
without using formal statistical methods, simply because the tools he
needed did not exist. Yet he shows a remarkable intuition for
statistical concepts, being quite aware of chance, variability and
random errors. This shows how Mendel was a man far ahead of his time.

Sir Ronald Fisher [1890--1962, Figure~\ref{Mendel}(b)] made
fundamental contributions to statistics and is also regarded as the
founder of quantitative genetics. He is described by Hald (\citeyear{Ha1998}) as
``\textit{a genius who almost single-handedly created the foundations
for modern statistical science}'' and by Dawkins (\citeyear{Da1995}) as ``\textit
{the greatest of Darwin's successors}.'' It is thus quite
understandable that Fisher became interested in Men\-del's work and data
very early in his career.

Let us now review the chronology of this controversy:
\begin{description}
\item[1856--1863] Mendel performed his experiments during this period.
He produced around 29,000 garden pea plants from controlled crosses and
registered several of their observable characteristics (phenotype),
such as shape and color of the seeds, height, flower color, etc.

\item[1865] Mendel presented the results of his experiments in a
communication entitled \textit{Experiments on Plant Hybridization}, read at two meetings
of the Society of Natural History of Br\"unn.

\item[1866] The paper with the same title was published in the
proceedings of that society. The paper had little impact and would be
cited only three times in the next 35 years.

\item[1900] His work was rediscovered independently by~Hu\-go de Vries,
Carl Correns and Erich von \mbox{Tschermak}.

\item[1902] The first statistical analysis of Mendel's data is
published in the first volume of Biometrika (Weldon, \citeyear{We1902}), using the
then recently invented chi-square test (Pearson, \citeyear{Pe1900}).

\item[1911] Fisher produced a first comment about Men\-del's results, in
a communication to the Cambridge University Eugenics Society, while he
was still an undergraduate:
``\textit{It is interesting
that Mendel's ori\-ginal results all fall within the limits of probable
\mbox{error}}'' and suggested that Mendel may have
``\textit{unconsciously
placed doubtful plants on the side which favoured his hypothesis}''
(Franklin et al., \citeyear{Fra2008}, page~16).

\item[1936] Fisher published the paper \textit{Has Mendel's work been
rediscovered?} (Fisher, \citeyear{Fi1936}), where he expresses the same concern but
this time presenting a detailed analysis, both of Mendel's experiments
and data. He also attributes the alleged forgery, not to Mendel
himself, but to an unknown assistant:
``\textit{Although no explanation can be expected
to be satisfactory, it remains a possibility among others that Mendel
was deceived by some assistant who knew too well what was expected}''
(Fisher, \citeyear{Fi1936}, page~132).
Fisher also questioned some other aspects of Mendel's experiments, but
those do not involve statistical aspects and will not be discussed here.

\item[1964] The publication De Beer (\citeyear{De1964}), intended to celebrate the
centennial of Mendel's article, highlights the fact that Fisher
``\textit{was able to reconstruct the sequence thread and development
of Mendel's series of experiments}'' and draws attention to Fi\-sher's
work on the statistical analysis of Mendel's results. Ironically,
Fisher's paper appears to have remained mostly overlooked until
approximately this anniversary, as far as we can tell based on the
scarcity of previous citations.

\item[1964--2007] During this period at least 50 papers~have been
published about the controversy created by Fisher. Some elucidative titles:
\textit{The too-good-to-be-true paradox and Gregor Mendel} (Pilgrim, \citeyear{Pi1984});
\textit{Are Mendel's results really too close?} (Edwards, \citeyear{Ed1986a});
\textit{Mud Sticks: On the Alleged Falsification of Mendel's
Data} (Hartl and Fairbanks,~\citeyear{HaFr2007}).

\item[2008] A group of scientists from different fields,\break  Franklin
(Physics and History of Science), Edwards (Biometry and Statistics, and
curiously, Fi\-sher's last student), Fairbanks (Plant and Wildlife
Sciences), Hartl (Biology and Genetics) and Seidenfeld (Philosophy and
Statistics), who have previously published work on the controversy,
merged their most relevant papers and published the book \textit{Ending
the Mendel--Fisher Controversy}. But is it really the end of the
controversy? The authors dismiss all of the issues raised by Fisher
except the ``\textit{too good to be true}'' (pages~68 and~310).

In a very interesting book review, entitled \textit{CSI: Mendel},
Stigler (\citeyear{Sti2008}) adds:
``\textit{\ldots  an actual end to that discussion is unlikely to be a
consequence of this book}.'' and
``\textit{\ldots  thanks to these lucid, insightful and balanced
articles, another generation will be able to join the quest with even
better understanding}.''
\end{description}

\section{Experiments and Data}\label{sec3}

Before introducing the data and discussing the corresponding
statistical analysis, it is important to understand the experiments and
the scientific hypotheses under evaluation.
Using a classification similar to that used by Fisher, the experiments
can be classified as follows: single trait, bifactorial, trifactorial and
gametic ratios experiments.

\begin{figure}

\includegraphics{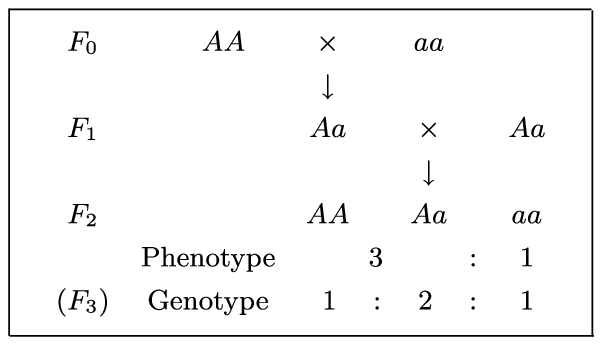}

\caption{A schematic representation of Mendel's single trait
experiments (in modern notation and terminology).}
\label{cross1}
\end{figure}

\textit{Single trait experiments}. These concern the transmission of
only one binary characteristic (or trait) at a time. Mendel examined
seven traits, two observable in the seeds (seed shape: round or
wrinkled; seed color: yellow or green) and five in the plants
(flower color: purple or white;
pod shape: inflated or constricted;
pod color: yellow or green;
flower position: axial or terminal;
stem length: long or short). First Mendel obtained what are now called
``pure lines,'' with each of the two forms of the seven characters,
that is, plants which yielded perfectly constant and similar offspring.
When crossing the two pure lines, $F_0$, for each character Mendel
observed that all the progeny, $F_1$, presented only one of the forms
of the trait. He called this one the \textit{dominant} form and
represented it by $A$. The other form was called \textit{recessive} and
denoted by $a$. In the seven traits listed above the first form is the
dominant and the second is the recessive. He then crossed the $F_1$
individuals (which he called the hybrids) and observed that in the
resulting generation, $F_2$, there were individuals of the two original
types, approximately in the ratio $3:1$ of the dominant type to the
recessive type.
In modern notation and terminology, we are studying a phenotype with
possible values ``$A$'' and ``$a$'' governed by a single gene with two
alleles ($A$ and $a$, where the first is dominant). The $F_0$ plants
are homozygous $AA$ (genotype $AA$, phenotype ``$A$'') or $aa$
(genotype $aa$, phenotype ``$a$''), the $F_1$ are all heterozygous $Aa$
(genotype $Aa$, phenotype ``$A$''), the $F_2$ plants can have genotype
$AA$ (phenotype ``$A$''), genotype $Aa$ (phenotype ``$A$'') and
genotype $aa$ (phenotype ``$a$'').
When Mendel self-fertilized the $F_2$ plants with phenotype ``$A,$'' he
found that about one-third of these always produced phenotype ``$A$''
progeny, while about two-thirds produced phenotype ``$A$'' and
phenotype ``$a$'' progeny in the ratio $3:1$.
This process is schematically represented in Figure~\ref{cross1},
where ($F_3$) refers to the progeny of the self-fertilized $F_2$ individuals.

\begin{table*}
\tabcolsep=0pt
\caption{Data given in Mendel (\protect\citeyear{Me1866}) for the single trait experiments.
``$A$'' (``$a$'') denotes the dominant (recessive) phenotype; $A$ ($a$)
denotes the dominant (recessive) allele; $n$ is the total number of
observations per experiment (that is, seeds for the seed trait
experiments and plants otherwise); $ n_{
{\fontsize{7.6pt}{9pt}\selectfont{\textnormal{``\textit{A}''}}}}$, $ n_{
{\fontsize{7.6pt}{9pt}\selectfont{\textnormal{``\textit{a}''}}}}$, $n_{Aa}$ and $n_{AA}$
denote observed frequencies}
\label{Tab1}
\begin{tabular*}{\textwidth}{@{\extracolsep{\fill}}llccd{4.0}d{4.0}d{4.0}c@{}}
\hline
& & & & & \multicolumn{2}{c}{\textbf{Obs. freq.}} & \multicolumn{1}{c@{}}
{\textbf{Theor. ratio}}
\\
 \cline{6-8}
& \multicolumn{1}{c}{\textbf{Trait}} & \multicolumn{1}{c}{\textbf{``\textit{A}''}} &
\multicolumn{1}{c}{\textbf{``\textit{a}''}} &
\multicolumn{1}{c}{$\bolds n$} & \multicolumn{1}{c}{$\bolds n_{
{\fontsize{7.6pt}{9pt}\selectfont{\textbf{``\textit{A}''}}}}$} &
\multicolumn{1}{c}{$\bolds n_{
{\fontsize{7.6pt}{9pt}\selectfont{\textbf{``\textit{a}''}}}}$} &
$\mbox{\textbf{``\textit{A}''}}\bolds:\mbox{\textbf{``\textit{a}''}}$ \\
\hline
& Seed shape & round & wrinkled & 7324 & 5474 & 1850 & $3:1$ \\
& Seed color & yellow & green & 8023 & 6022 & 2001 & $3:1$ \\
& Flower color & purple & white & 929 & 705 & 224 & $3:1$ \\
$F_2$ & Pod shape & inflated & constricted & 1181 & 882 & 299 & $3:1$
\\
& Pod color & yellow & green & 580 & 428 & 152 & $3:1$ \\
& Flower position & axial & terminal & 858 & 651 & 207 & $3:1$ \\
& Stem length & long & short & 1064 & 787 & 277 & $3:1$ \\
 [6pt]
& \multicolumn{1}{c}{\textbf{Trait}} & \multicolumn{1}{c}{$\bolds A$} &
\multicolumn{1}{c}{$\bolds a$} &
\multicolumn{1}{c}{$\bolds n$} & \multicolumn{1}{c}{$\bolds{n_{Aa}}$} &
\multicolumn{1}{r}{$\bolds {n_{AA}}$} & $\bolds {Aa:AA}$ \\
& Seed shape & round & wrinkled & 565 & 372 & 193 & $2:1$ \\
& Seed color & yellow & green & 519 & 353 & 166 & $2:1$ \\
& Flower color & purple & white & 100 & 64 & 36 & $2:1$ \\
$(F_3)$ & Pod shape & inflated & constricted & 100 & 71 & 29 & $2:1$ \\
& Pod color & yellow & green & 100 & 60 & 40 & $2:1$ \\
& Flower position & axial & terminal & 100 & 67 & 33 & $2:1$ \\
& Stem length & long & short & 100 & 72 & 28 & $2:1$ \\
& Pod color (rep.) & yellow & green & 100 & 65 & 35 & $2:1$ \\
\hline
\end{tabular*}
\end{table*}

Table~\ref{Tab1} presents the data given in Mendel (\citeyear{Me1866}) for the
single trait experiments just described.
As an illustration of the variability of the results between plants,
Mendel also presented the individual figures obtained for the ten first
plants of each of the experiments relative to the seed characteristics
(these are referred to by Fisher as ``\textit{illustrations of plant
variation},'' cf. Table~\ref{TableV}).

\begin{table}
\tabcolsep=0pt
\caption{Data from the bifactorial experiment [as organized by Fisher  (\protect\citeyear{Fi1936})]}
\label{Tab2}
\begin{tabular*}{\columnwidth}{@{\extracolsep{\fill}}lccc@{\hspace*{3pt}}c@{}}
\hline
& $\bolds{AA}$ & $\bolds{Aa}$ & $\bolds{aa}$ &  \textbf{Total}  \\
\hline
$BB$ & \hphantom{1}38 & \hphantom{1}60 & \hphantom{1}28 & 126 \\
$Bb$ & \hphantom{1}65 & 138 & \hphantom{1}68 & 271 \\
$bb$ & \hphantom{1}35 & \hphantom{1}67 & \hphantom{1}30 & 132 \\
 [3pt]
 Total  & 138 & 265 & 126 & 529 \\
\hline
\end{tabular*}
\end{table}

\begin{figure}

\includegraphics{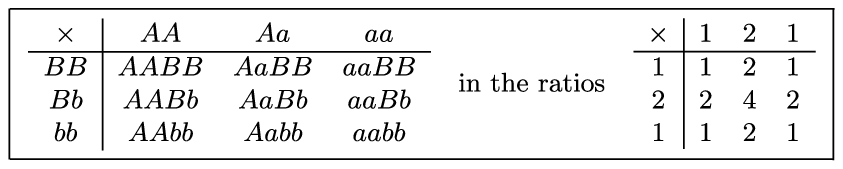}

\caption{Genotypes and theoretical ratios for the bifactorial experiment.}
\label{Fig3}
\end{figure}

\textit{Bifactorial experiment}. This is an experiment similar to the
single trait experiments but observing two characteristics
simultaneously (seed shape, A, and seed color, B, starting from pure
lines on both). The aim was to observe how the two traits are combined.
Mendel postulated and confirmed from the results of the experiment that
the traits considered are assorted independently.\footnote{This
independence hypothesis is also a matter of controversy (did Mendel
detect linkage?) and has been discussed thoroughly in the literature
(see Franklin et al., \citeyear{Fra2008}, pages 288--292).} That is, given a
trait A with an $F_2$ generation $AA$, $Aa$ and $aa$ in the ratio
$1:2:1$, and a trait B with $BB$, $Bb$ and $bb$ in the same ratio,
combining the two independently leads to the genotypes and theoretical
ratios represented in Figure~\ref{Fig3}. The data, organized by Fisher
from Mendel's description, are shown in Table~\ref{Tab2}.

\begin{table*}[t]
\tabcolsep=0pt
\caption{Data from the trifactorial experiment [as organized by
Fisher  (\protect\citeyear{Fi1936})]}
\label{Tab3}
\begin{tabular*}{\textwidth}{@{\extracolsep{\fill}}lcccc@{\hspace*{3pt}}cccc@{\hspace*{3pt}}
cccc@{\hspace*{3pt}}cccc@{}}
\hline
& \multicolumn{4}{c@{\hspace*{3pt}}}{$\bolds{CC}$} & \multicolumn{4}{c@{\hspace*{3pt}}}
{$\bolds{Cc}$} & \multicolumn{4}{c@{\hspace*{3pt}}}{$\bolds{cc}$} &
\multicolumn{4}{c@{}}{\textbf{Total}} \\
\ccline{2-5,6-9,10-13,14-17}
& $\bolds{AA}$ & $\bolds{Aa}$ & $\bolds{aa}$ &  \textbf{Total}  & $\bolds{AA}$ &
$\bolds{Aa}$ & $\bolds{aa}$ &  \textbf{Total}  & $\bolds{AA}$ & $\bolds{Aa}$
& $\bolds{aa}$ &  \textbf{Total}  & $\bolds{AA}$ & $\bolds{Aa}$ & $\bolds{aa}$ &
\textbf{Total}  \\
\hline
$BB$ & \hphantom{1}8 & 14 & \hphantom{1}8 & \hphantom{1}30 & 22 & \hphantom{1}38 & 25 &
\hphantom{1}85 & 14 & 18 & 10 & \hphantom{1}42 & \hphantom{1}44 & \hphantom{1}70&
\hphantom{1}43 & 157 \\
$Bb$ & 15 & 49 & 19 & \hphantom{1}83 & 45 & \hphantom{1}78 & 36 & 159 & 18 & 48 & 24 &
\hphantom{1}90 & \hphantom{1}78 &175 & \hphantom{1}79 & 332 \\
$bb$ & \hphantom{1}9 & 20 & 10 & \hphantom{1}39 & 17 & \hphantom{1}40 & 20 &
\hphantom{1}77 & 11 & 16 & \hphantom{4}7 & \hphantom{1}34 & \hphantom{1}37 & \hphantom{1}76&
\hphantom{1}37 & 150 \\
[3pt]
 Total  & 32 & 83 & 37 & 152 & 84 & 156 & 81 & 321 & 43 & 82 & 41& 166 & 159 & 321 & 159 & 639 \\
\hline
\end{tabular*}
\end{table*}

\begin{figure}[t]

\includegraphics{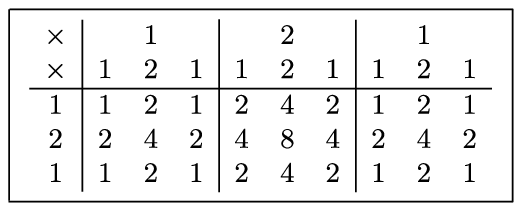}

\caption{Theoretical ratios for the trifactorial experiment.}
\label{Fig4}

\end{figure}

\textit{Trifactorial experiment}. This experiment is also similar to
the previous experiment but considering the crossing of three traits
(seed shape, seed color and flower color). The data, organized by
Fisher from Mendel's description, are shown in Table~\ref{Tab3},\vadjust{\goodbreak}
whereas the corresponding theoretical ratios are gi\-ven in Figure~\ref{Fig4}.

\begin{figure*}[b]
\centering
\begin{tabular}{c}

\includegraphics{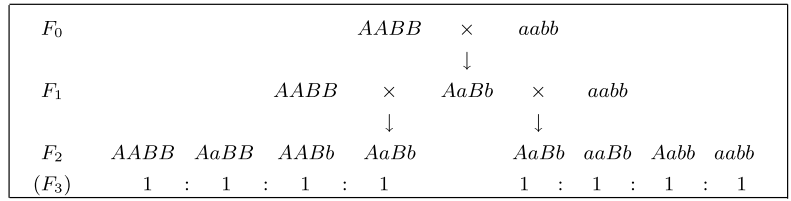}
\\
\footnotesize{(a)}\\

\includegraphics{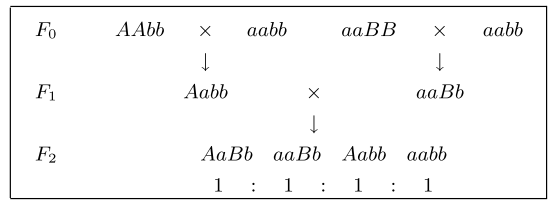}
\\
\footnotesize{(b)}
\end{tabular}
\caption{Schematic representation of the gametic ratios experiments:
\textup{(a)} experiments 1--4 in Table~\protect\ref{Tab4}; \textup{(b)} experiment 5 in
Table~\protect\ref{Tab4}.}
\label{Fig5}
\end{figure*}

\textit{Gametic ratios experiments}. In this last series of
experiments Mendel designed more elaborated crosses in order to obtain
``\textit{conclusions as regards the composition of the egg and pollen
cells of hybrids}.'' The crosses are represented in Figure~\ref{Fig5}
and the data are shown in Table~\ref{Tab4}.

We will also use an organization of the data into 84 binomial
experiments, similar to the one proposed by Edwards (\citeyear{Ed1986a}, see also
Franklin et al., \citeyear{Fra2008}, Chapter~4). The data set used is
described in detail in Appendix~\ref{apC}.

All the computations and Monte Carlo simulations described were carried
out using the \texttt{R} software (R Development Core Team, \citeyear{R2008}).
The full code is available upon request.

\section{Statistical Analysis: Incriminating Evidence}\label{sec4}

\subsection{Fisher's Analysis}\label{sec4.1}
As mentioned in Section~\ref{sec2}, Fisher (\citeyear{Fi1936}) presents a very detailed
analysis of both Mendel's experiments and data. Here we will
concentrate on a particular part of the analysis, the chi-square
analysis summarized in Table V, page~131, of Fisher (\citeyear{Fi1936}), which is
reproduced in Table~\ref{TableV}.
This table has been the subject of a lot of debate, and it constitutes
the main evidence for the ``\textit{too good to be true}'' aspect of
Mendel's data as claimed by Fisher. We later present a new explanation
for this evidence.

\begin{table}
\tabcolsep=0pt
\caption{Data from the gametic ratios experiments (Mendel,\protect\citeyear{Me1866})}
\label{Tab4}
\begin{tabular*}{\columnwidth}{@{\extracolsep{4in minus 4in}}lc@{\hspace*{3pt}}ccccccc@{}}
\hline
\multirow{2}{18pt}{\textbf{Exp.}}
&\multicolumn{1}{c@{\hspace*{3pt}}}{\multirow{2}{5pt}{$\bolds n$}}&
\multicolumn{4}{c}{\multirow{2}{46pt}{\centering\textbf{Observed frequencies}}}
& \multirow{2}{44pt}{\centering\textbf{Theoretical ratio}}&
\multicolumn{2}{c@{}}{\textbf{Traits}}\\
\cline{8-9}
&&&&&& &$\bolds A$ &  $\bolds B$ \\
\hline
1 & \hphantom{1}90 & 20 & 23 & 25 & 22 & $1:1:1:1$ & seed shape & seed color \\
2 & 110 & 31 & 26 & 27 & 26 & $1:1:1:1$ & seed shape & seed color \\
3 & \hphantom{1}87 & 25 & 19 & 22 & 21 & $1:1:1:1$ & seed shape & seed color \\
4 & \hphantom{1}98 & 24 & 25 & 22 & 27 & $1:1:1:1$ & seed shape & seed color \\
5 & 166 & 47 & 40 & 38 & 41 & $1:1:1:1$ & flower color & stem length \\
\hline
\end{tabular*}
\end{table}

The analysis is very simple to describe: for each separate experiment,
Fisher performed a chi-square goodness-of-fit test, where $H_0$
specifies the probabilities implied by the theoretical ratios. Note
that, for the two category cases, this is equivalent to the usual
asymptotic test for a single proportion. Then he aggregated all the
tests by summing the chi-square statistics as well as the associated
number of degrees of freedom and computed an aggregated $p$-value of
0.99993. This would mean that if Mendel's experiments were repeated,
under ideal conditions such that all the null-hypotheses are true, and
all the Bernoulli trials---within and between experiments---are
independent, the probability of getting an overall better result would
be 7$/$100,000.

Fisher's chi-square results were recomputed just to confirm that we are
working with exactly the same data and assumptions. The results, given
in the first 4 columns of Table~\ref{Tab5}, show that the statistics
($\chi^2_{\mathrm{obs}}$) are identical to Fisher's values, but there are some
differences in the $p$-values which certainly reflect different methods
of computing the chi-square distribution function [$p$-value ($\chi
^2_{\mathrm{df}}$) denotes the \mbox{$p$-value} computed from a $\chi^2$ distribution
with $\mathit{df}$ degrees of freedom, that is, $P( \chi^2_{\mathrm{df}} >
\chi^2_{\mathrm{obs}} )$].

\begin{table}
\tabcolsep=0pt
\caption{Fisher's chi-square analysis (``\textit{Deviations expected
and observed in all experiments}'')}
\label{TableV}
\begin{tabular*}{\columnwidth}{@{\extracolsep{\fill}}ld{2.0}d{2.4}d{1.5}@{}}
\hline
\multirow{4}{50pt}{\textbf{Experiments}}
&  \multicolumn{1}{c}{\multirow{4}{46pt}{\textbf{Expectation}}}
& \multicolumn{1}{c}{\multirow{4}{11pt}{$\bolds{\chi^2}$}}
&  \multicolumn{1}{c@{}}{\textbf{Probability}} \\
&&&\multicolumn{1}{c@{}}{\textbf{of exceeding}} \\
&&&\multicolumn{1}{c@{}}{\textbf{deviations}} \\
&&&\multicolumn{1}{c@{}}{\textbf{observed}}\\
\hline
$3:1$ ratios & 7    & 2.1389 &  0.95 \\
$2:1$ ratios & 8    & 5.1733 &  0.74 \\
Bifactorial & 8    & 2.8110 &  0.94 \\
Gametic ratios & 15    & 3.6730 &  0.9987 \\
Trifactorial & 26    & 15.3224 &  0.95 \\
[3pt]
  Total & 64    & 29.1186 &  0.99987 \\
Illustrations of \\ \quad plant variation & 20    & 12.4870 &
0.90 \\
 [3pt]
  Total & 84    & 41.6056 &  0.99993 \\
\hline
\end{tabular*}
\end{table}

The table also gives the Monte Carlo (MC) estimates of the $p$-values
(and corresponding standard errors, \textit{se}) based on 1,000,000 random
repetitions of the experiments using binomial or multinomial sampling,
whichever is appropriate, as considered~by~Fi\-sher. A~more detailed
description of the Monte Carlo simulation is given in Appendix~\ref{apA1}.

Comparing the list of $\chi^2_{\mathrm{df}}$ $p$-values with the list of MC $p$-values,
the conclusion is that the approximation of the sampling
distribution of the test statistic
by the chi-square distribution is very accurate, and that Fisher's
analysis is very solid [our results are also in accordance with the
results of similar but less extensive simulations described in
Novitski  (\citeyear{Novi1995})].
Moreover, we can also conclude
that the evidence ``against'' Mendel is greater than that given by
Fisher, since an estimate of the probability of getting an overall
better result is now 2$/$100,000.
We have also repeated the chi-square analysis considering the 84
binomial experiments
(results given in the next 3 columns of Table~\ref{Tab5}) and concluded
that the two sampling models are almost equivalent, the second one
(only binomial) being slightly more favorable to Fisher and less
favorable to Mendel. Acting in Mendel's defense, the results will be
more convincing if we prove our case under the least favorable
scenario. Thus, for the remaining investigation, we use only the
binomial model and data set.

\begin{sidewaystable*}
\tabcolsep=0pt
\tablewidth=\textwidth
\caption{Results of the chi-square analysis considering different
models and methods for computing/estimating $p$-values. Each line
corresponds to a different type of experiment in Mendel's paper: single
trait, $3: 1$ ratios; single trait, $2: 1$ ratios; bifactorial (BF); gametic
ratios (GR); trifactorial (TF); and illustrations of plant variation
(PV). $\mathit{df}$: degrees of freedom of the asymptotic distribution of the
$\chi^2$ test statistic under $H_0$ (Mendel's theory); $\chi^2_{\mathrm{obs}}$:
observed value of the $\chi^2$ test statistic; $p$-value ($\chi^2_{\mathrm{df}}$):
$p$-value computed assuming that the test statistic follows, under $H_0$,
a $\chi^2$ distribution with $\mathit{df}$ degrees of freedom; $p$-value
(MC):
$p$-value estimated from Monte Carlo simulation. \textit{se}: standard error of the
$p$-value (MC) estimate}
\label{Tab5}
\begin{tabular*}{\textwidth}{@{\extracolsep{\fill}}lcd{2.4}d{2.5}d{2.7}d{2.4}
d{2.5}d{2.7}d{2.6}d{2.5}d{2.5}d{2.6}d{2.5}d{2.5}@{}}
\hline
&& \multicolumn{3}{c}{\multirow{2}{99pt}[-3pt]{\centering\textbf{Fisher
\mbox{(binomial $\bolds +$ multinomial)}}}} &   &&&
\multicolumn{3}{c}{\textbf{Model A}} & \multicolumn{3}{c@{}}{\textbf{Model B}} \\
\ccline{9-11,12-14}
&&& &  & \multicolumn
{3}{c}{\textbf{Edwards (binomial)}} & \multicolumn{1}{c}{$\bolds{\alpha=0.094}$}
& \multicolumn{1}{c}{$\bolds{\alpha=0.201}$} & \multicolumn
{1}{c}{$\bolds{\alpha=0.362}$} & \multicolumn{1}{c}{$\bolds{\beta=0.261}$} &
\multicolumn{1}{c}{$\bolds{\beta=0.455}$} &
\multicolumn{1}{c@{}}{$\bolds{\beta=0.634}$} \\
\ccline{3-5,6-8,9-11,12-14}
&& & \multicolumn{1}{c}{$\bolds p$\textbf{-value}} &
\multicolumn{1}{c}{$\bolds p$\textbf{-value}} & &
\multicolumn{1}{c}{$\bolds p$\textbf{-value}} &
\multicolumn{1}{c}{$\bolds p$\textbf{-value}} &
\multicolumn{1}{c}{$\bolds p$\textbf{-value}} &
\multicolumn{1}{c}{$\bolds p$\textbf{-value}} &
\multicolumn{1}{c}{$\bolds p$\textbf{-value}} &
\multicolumn{1}{c}{$\bolds p$\textbf{-value}} &
\multicolumn{1}{c}{$\bolds p$\textbf{-value}} &
\multicolumn{1}{c}{$\bolds p$\textbf{-value}} \\
\textbf{Exp.} & $\bolds{df}$ & \multicolumn{1}{c}{$\bolds{\chi^2_{\mathbf{obs}}}$} &
\multicolumn{1}{c}{\textbf{($\bolds{\chi^2_{\mathbf{df}}}$)}} &
\multicolumn{1}{c}{\textbf{(MC)}} &
\multicolumn{1}{c}{$\bolds{\chi^2_{\mathbf{obs}}}$} &
\multicolumn{1}{c}{\textbf{($\bolds{\chi^2_{\mathbf{df}}}$)}} &
\multicolumn{1}{c}{\textbf{(MC)}} & \multicolumn{1}{c}{\textbf{(MC)}} &
\multicolumn{1}{c}{\textbf{(MC)}} & \multicolumn{1}{c}{\textbf{(MC)}} &
\multicolumn{1}{c}{\textbf{(MC)}} & \multicolumn{1}{c}{\textbf{(MC)}} &
\multicolumn{1}{c}{\textbf{(MC)}} \\
& & & & \multicolumn{1}{c}{\textbf{(\textit{se})}} && &
\multicolumn{1}{c}{\textbf{(\textit{se})}}
& \multicolumn{1}{c}{\textbf{(\textit{se})}} & \multicolumn{1}{c}{\textbf{(\textit{se})}} &
\multicolumn{1}{c}{\textbf{(\textit{se})}} & \multicolumn{1}{c}{\textbf{(\textit{se})}} &
\multicolumn
{1}{c}{\textbf{(\textit{se})}} & \multicolumn{1}{c}{\textbf{(\textit{se})}} \\
\hline\\
3:1 & \hphantom{6}7 & 2.1389 & 0.9518 & 0.9519 & 2.1389 & 0.9518 & 0.9517 &
0.9069 & 0.8286 & 0.6579 & 0.9023 & 0.8446 & 0.7701 \\
& & & & (0.0002) && & (0.0002) &
(0.0003) & (0.0004) & (0.0005)
& (0.0003) & (0.0004) &
(0.0004) \\
2:1 & \hphantom{6}8 & 5.1733 & 0.7389 & 0.7401 & 5.1733 & 0.7389 & 0.7393 &
0.4955 & 0.2374 & 0.1156 & 0.6044 & 0.4826 & 0.3586 \\
& & & & (0.0004) && & (0.0004) &
(0.0005) & (0.0004) & (0.0003)
& (0.0005) & (0.0005) &
(0.0005) \\
BF & \hphantom{6}8 & 2.8110 & 0.9457 & 0.9462 & 2.7778 & 0.9475 & 0.9482 & 0.8838
& 0.7839 & 0.5926 & 0.8914 & 0.8248 & 0.7376 \\
& & & & (0.0002) && & (0.0002) &
(0.0003) & (0.0004) & (0.0005)
& (0.0003) & (0.0004) &
(0.0004) \\
GR & 15 & 3.6730 & 0.9986 & 0.9987 & 3.6277 & 0.9987 & 0.9987 &
0.9950 & 0.9811 & 0.9063 & 0.9939 & 0.9827 & 0.9584 \\
& & & & (0.00004) && & (0.00004) &
(0.00007) & (0.0001) & (0.0003)
& (0.00008) & (0.0001) &
(0.0002) \\
TF & 26 & 15.3224 & 0.9511 & 0.9512 & 15.1329 & 0.9549 & 0.9555 &
0.6973 & 0.2917 & 0.0812 & 0.8493 & 0.6941 & 0.4847 \\
& & & & (0.0002) && & (0.0002) &
(0.0004) & (0.0005) & (0.0003)
& (0.0004) & (0.0005) &
(0.0005) \\
[6pt]
Tot. & 64 & 29.1185 & 0.99995 & 0.99995& 28.8506 & 0.99995 & 0.99995
& 0.9917 & 0.8175 & 0.2965 & 0.9980 & 0.9800 & 0.8887 \\
& & & & (0.000007) && & (0.000007) &
(0.00008) & (0.0004) & (0.0005)
& (0.00004) & (0.0001) &
(0.0003) \\
PV & 20 & 12.4870 & 0.8983 & 0.9000& 12.4870 & 0.8983 & 0.9003 &
0.5932 & 0.2196 & 0.0684 & 0.7582 & 0.5922 & 0.4028 \\
& & & & (0.0003) && & (0.0003) &
(0.0005) & (0.0004) & (0.0003)
& (0.0004) & (0.0005) &
(0.0005) \\
 [6pt]
Tot. & 84 & 41.6055 & 0.99997 & 0.99998 & 41.3376 & 0.99998 & 0.99998
& 0.9860 & 0.6577 & 0.1176 & 0.9980 & 0.9733 & 0.8348 \\
& & & & (0.000005) && & (0.000005) &
(0.0001) & (0.0005) & (0.0003)
& (0.00004) & (0.0002) &
(0.0004) \\
\hline
\end{tabular*}
\end{sidewaystable*}

Franklin et al. (\citeyear{Fra2008}, pages~29--67) provide a comprehensive
systematic review of all the papers published since the 1960s in
reaction to Fisher's accusations. The vast majority of those authors
try to put \mbox{forward} arguments in Mendel's defense.
We only highlight here some of the more relevant contributions
regarding specifically the ``\textit{too good to be true}'' conclusion
obtained from the chi-square analysis.
The majority of the reactions/arguments can be generically \mbox{classified}
into three categories.

In the first category we consider those who do not believe in Fisher's
analysis. This is the case of  Pilgrim (\citeyear{Pi1984,Pi1986}) who in the first
paper affirms to have detected ``\textit{four paradoxical elements in
Fisher's reasoning}'' and who, in the second, claims to have been able
to show where Fisher went wrong. Pilgrim's arguments are related to the
application of the chi-square global statistic and were refuted by
Edwards (\citeyear{Ed1986b}).

As a second category, those who, in spite of believing that Fisher's
analysis is correct, think it is too demanding and propose alternative
ways to analyze Mendel's data.
Edwards (\citeyear{Ed1986a}) analyzes the distribution of a set of test statistics, whereas
Seidenfeld (\citeyear{Sei1998}) analyzes the distribution of a set of $p$-values. They
both find the ``\textit{too good to be true}'' characteristic and come
to the conclusion that Mendel's results were adjusted rather than
censored.\footnote{These are the precise words used in the cited
references (Edwards, \citeyear{Ed1986b}  and Seidenfeld, \citeyear{Sei1998}). They mean that some
results have been slightly modified to fit Mendel's expectations
(``adjusted''), instead of just being eliminated
(``censored'' or ``truncated'').}
The methods of Leonard (\citeyear{Leo1977}) and Robertson (\citeyear{Ro1978}), who analyzed
only a small part of the data, could also be classified here, but, according
to Piegorsch (\citeyear{Pie1983}), their contribution to advance the debate was marginal.

Finally, as a third category,
those who believe\break  Fisher's analysis is correct under its assumptions
(binomial/multinomial sampling, independent experiments) and then try
to find a reason or explanation, other than deliberate cheating, for
the observation of a very high global $p$-value. Such an explanation has
to imply the failure of at least one of those two assumptions.
Moreover, that failure has to occur in a specific direction, the one
which would reduce the chi-square statistics: for instance, the
distribution of the phenotypes is not binomial and has a smaller
variance than the binomial.
The various explanations that have been put forward can be divided
into the following: biological, statistical and metho\-dological.

Among the biological candidate explanations, one that received some
attention was the ``Tetrad Pollen Model'' (see Fairbanks and Schaalje, \citeyear{Fai2007}).

Few purely statistical explanations have been proposed and most of them
are anecdotal. One that raised some discussions was a proposal of
Weiling (\citeyear{Wei1986}) who considers, based on the tetrad pollen model just
mentioned, a distribution with smaller variance than the binomial for
some of the experiments, and hypergeometric for other experiments.

The majority of the suggested explanations are of a methodological
nature: the ``anonymous'' assistant (Fisher); sequential procedures,
like stopping the count when the results look good (several authors);
discard plants or complete experiments due to suspicions of some
experimental error, like pollen contamination (Dobzhansky, \citeyear{Dob1967});
luck(?); inherent difficulties in the classification of the phenotypes
(Root-Bernstein, \citeyear{Ro1983}); data selection for presentation
(Di Trocchio, \citeyear{DiT1996};
Fairbanks and Rytting,~\citeyear{Fai2001}).

It is important to keep in mind that for an explanation to be
acceptable as the solution to the controversy it must fulfill a number
of conditions: (i) it must be biologically plausible and/or
experimentally verifiable;
(ii) it must be statistically correct and pass the chi-square and
eventually other statistical analyses aiming at disentangling the
enigma; and (iii) assuming that Mendel's theory is correct and that he
is not guilty of any deliberate fraud, it has to find support in and it
can not contradict Mendel's writings.
The fact is that all the explanations which were proposed up to now
failed in one or other of these requirements.

In summary, Fisher's analysis has resisted all attempts to be either
refuted or explained. Our simulation also confirms that, under the
standard assumptions, Fisher's tests and conclusions are correct.

\subsection{Analysis of $p$-Values}\label{sec4.2}

As mentioned in Section~\ref{sec3}, Edwards (\citeyear{Ed1986a}) proposed an organization of
the data into 84 binomial experiments. He then used the data to compute
what he called (signed) $\chi$ values,
that is, the square root of the chi-square statistic with the sign of
the deviation (``$+$'' if observed $>$ expected and ``$-$'' if observed
$<$ expected). Since all the tests have one degree of freedom and,
assuming that Mendel's theory is correct, the $\chi$ values should
follow approximately a standard normal distribution. However, a normal
qqplot of those values shows apparently a large deviation from
normality (Franklin et al., \citeyear{Fra2008}, Figures~1.1 and~1.2,
page~49). From the shape of the plot Edwards (\citeyear{Ed1986a}) concluded that it
appears to be more likely that Mendel's results were adjusted rather
than truncated. This conclusion, to which Seidenfeld (\citeyear{Sei1998}) also
arrives, and later Franklin et al. (\citeyear{Fra2008}) agree, would render
some of the most plausible methodological explanations not viable.

Another approach is to analyze
the $p$-values of the individual $\chi^2_1$ tests. This idea was explored
by Seidenfeld (\citeyear{Sei1998}); (see also Franklin et al., \citeyear{Fra2008},
Figures~1.3 and 1.4, page~59), although not so systematically as in the analysis
provided here. The 84 $\chi$ values, along with the 84 $p$-values, are
also given in Appendix~\ref{apC}.

\begin{figure}

\includegraphics{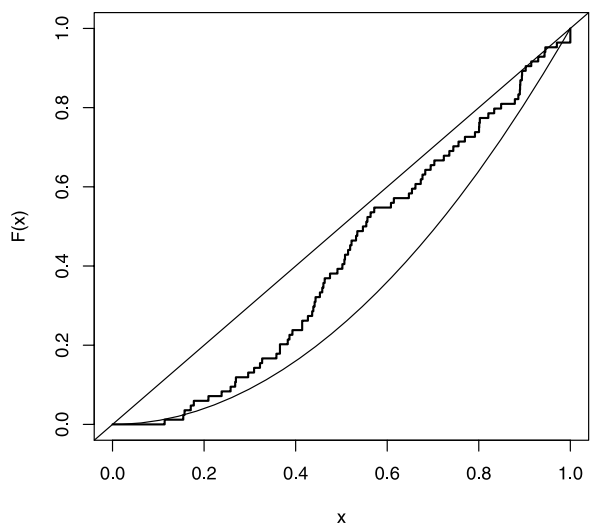}

\caption{Empirical cumulative distribution function of the $p$-values
(stair steps line); cumulative distribution function of the uniform
$(0,1)$ random variable (straight line); cumulative distribution function
of the maximum of two $(0,1)$ uniform random variables (curve).}
\label{pval1}
\end{figure}

As for Fisher's and Edwards' analysis, we know what to expect under the
ideal assumptions. That is, if: (i) Mendel's theory is valid for all
the experiments, or, equivalently, if the null hypotheses of the
chi-square tests are true in all cases; (ii) the experiments were
performed independently and as described in Mendel's paper; and (iii)
the chi-square approximation is valid, then the $p$-values follow a
uniform $(0,1)$ distribution. Therefore, the plot of the empirical
cumulative distribution function (e.c.d.f.)\footnote{The e.c.d.f. is defined,
for a random sample of size $n$, as $F_n(x)= \{\mbox{n. of
observations} \le x\}/n$.} of the $p$-values should be close to the
diagonal of the $(0,1)\times(0,1)$ square. However, the e.c.d.f., plotted
in Figure~\ref{pval1}, reveals a marked difference from uniformity.
This visual assertion
was confirmed by a Kolmogorov--Smirnov (K--S) goodness-of-fit test
($p$-value${}=0.0036$, details in Appendix~\ref{apA2}). We can therefore
conclude with a high confidence that the distribution of the $p$-values
deviates from a uniform $(0,1)$ distribution. It is then natural to
wonder about the kind of deviation and its meaning. In Figure~\ref
{pval1} we also plot the cumulative distribution function (c.d.f.) of the
maximum of two uniform $(0,1)$ random variables, $y=x^2$, since this is
central to the explanation that we give later for Mendel's results.

\begin{figure}

\includegraphics{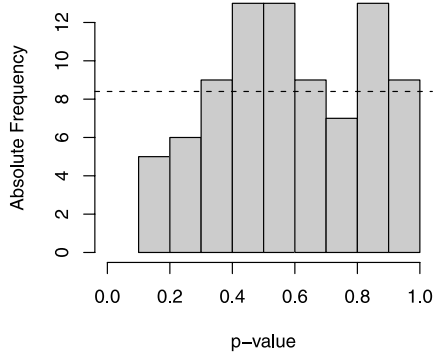}

\caption{Histogram of the 84 $p$-values observed [the dashed line
indicates the expected frequencies under the uniform $(0,1)$ distribution].}
\label{hist1}
\end{figure}

The histogram of the $p$-values (Figure~\ref{hist1}) is helpful for our
argumentation. One could perhaps think that the uniform distribution is
not a good fit for the sample of $p$-values because some of the null
hypotheses are not true. But if that were the case, we would observe an
excess of values close to 0, and the histogram shows precisely the
opposite. Possible reasons for this to happen are as follows: either
the data shows that the hypotheses are ``more true,'' that is, the data
are better than expected under the null hypotheses, or there is
something wrong with the assumptions (and possible explanations are, as
for the chi-square analysis, smaller variance than binomial, or lack of
independence).

In conclusion, the $p$-value analysis shows that the probability of
obtaining overall results as good or better than those obtained by
Mendel (under the assumptions) is about 4$/$1000. This ``evidence'' is
not as extreme as the 2$/$100,000 resulting from the chi-square analysis
but points in the same direction.

\section{A Plausible Explanation}\label{sec5}
\subsection{A Statistical Model for the $p$-Values}\label{sec5.1}

In the previous section we have shown that there is strong evidence
that the $p$-values are not uniformly distributed. What is then their
distribution, and how can it be explained?

The shape of the e.c.d.f. provides a hint: it resembles the function $x^2$,
which is the c.d.f. of the maximum of two uniform $(0,1)$ random variables,
as can easily be shown.
It appears that some c.d.f. intermediate between that corresponding to a
uniform $(0,1)$ random variable and that corresponding to the maximum of
two uniform $(0,1)$ random variables best fits the e.c.d.f. of the sample
$p$-values (see Figure~\ref{pval1}).

One explanation for this is the following: suppose Mendel has repeated
some experiments, presumably those which deviate most from his theory,
and reports only the best of the two. A related possibility was
suggested by Fairbanks and Rytting (\citeyear{Fai2001}, page~743): ``\textit{We
believe that the most likely explanation of the bias in
Mendel's data is also the simplest. If Mendel selected for presentation
a subset of his experiments that best represented his
theories, $\chi^2$ analysis of those experiments should display a
bias}.'' The authors support this explanation with citations from
Mendel's work. We have found only an attempt to verify the effect of
such a selection procedure on the chi-square analysis (footnote
number~62, page~73, of Franklin et al., \citeyear{Fra2008}), but it seems to
lead to the wrong conclusion, as we conclude later that the effect of
the given explanation on the chi-square analysis is very small.
Moreover, the explanation appears to have been abandoned because
``\textit{It does not [however] address the demonstration, by both
Edwards and Seidenfeld, that Mendel's data had not merely been
truncated, but adjusted}'' (Franklin et al., \citeyear{Fra2008}, page~62).

Both procedures described in the previous paragraph for selecting the
data to be presented can be modeled by assuming that an experiment is repeated
whenever its $p$-value is smaller than $\alpha$, where $0 \le\alpha\le
1$ is a parameter fixed by the experimenter, and then only the one with
the largest $p$-value is reported.\footnote{Note that this is just an
idealized model on which to base our explanation. We are not suggesting
that Mendel actually computed $p$-values!}
Under this selection model (from now on named ``model A''), the c.d.f. of
the $p$-values of the experiments reported
is given by
%
\begin{equation}
\label{FmodelA} \qquad
F_\alpha(x) =
\cases{\displaystyle
x^2  ,&   if   $0 \le x \le\alpha $,\cr\displaystyle
(1 + \alpha)x - \alpha
  ,&   if   $\alpha< x \le1$.
}
\end{equation}

\begin{pf} For a given experiment, denote by $X$ the $p$-value
effectively reported.
We have that $X=X_1$, if $X_1 \ge\alpha$ and $X=\max(X_1,X_2)$ if
$X_1 < \alpha$, where $X_1$ and $X_2$ represent the $p$-values obtained
in the first and the second realization of the experiment (if there is
one), respectively. Assume that $X_1$ and $X_2$ are independent and
identically distributed continuous uniform $(0,1)$ random variables (i.e.,
the two realizations of the experiment are independent and the
associated null hypothesis is true), that is, $P(X_1\le x)=P(X_2\le
x)=x $, $0\le x \le1$. In the derivation of $F_\alpha(x)=P(X \le x)$,
the cases $0 \le x < \alpha$ and $\alpha\le x \le1$ are considered
separately.

If $0 \le x < \alpha$,
\begin{eqnarray*}
P ( X \le x  )
& = & P \bigl( \max(X_1,X_2) \le x  \bigr)\\
 &= &P
 ( \{X_1\le x\} \cap\{X_2\le x \}  )  \\
 & = & P
( X_1 \le x  ) P ( X_2 \le x  ) = x^2.
\end{eqnarray*}

If $ \alpha\le x \le1$,
\begin{eqnarray*}
&&P ( X \le x  ) \\
     && \quad =      P \bigl(  ( \{X \le x\}
\cap\{X_1 < \alpha\}  )\\
&& \qquad \hphantom{P} {}\cup ( \{X \le x\} \cap\{X_1 \ge
\alpha\}  )  \bigr)
\\
 && \quad =  P  ( \{X \le x\} \cap\{X_1 < \alpha\}  )\\
&& \qquad {}+ P  ( \{X \le x\} \cap\{X_1 \ge\alpha\}  )  \\
 && \quad =  P \bigl ( \{\max(X_1,X_2) \le x\} \cap\{X_1 <
\alpha\}  \bigr)\\
&& \qquad {}+ P  ( \{X_1 \le x\} \cap\{X_1 \ge\alpha\}
)
\\
 && \quad =     P  ( \{X_1 < \alpha\} \cap\{X_2 \le x\}
) + P  ( \alpha\le X_1 \le x  )\\
 && \quad = x \times\alpha+ (x -
\alpha).
\end{eqnarray*}
\upqed
\end{pf}

Suppose that model A holds but $\alpha$ is unknown and must be
estimated using the available sample of 84 binomial $p$-values. The
minimum distance estimator based on the Kolmogorov distance, also
called the ``Minimum Kolmogorov--Smirnov test statistic
estimator'' (Easterling, \citeyear{Eas1976}), provides one method for estimating
$\alpha$. This estimate is the value of $\alpha$ which minimizes the
K--S statistic,
%
\begin{equation}
D(\alpha)=\sup_x  | F_n(x) - F_\alpha(x)  |
\label{eqDalfa}
\end{equation}
for testing the null hypothesis that the c.d.f. of the $p$-values is
$F_\alpha$. Equivalently, the estimate can be determined by finding the
value of $\alpha$ which maximizes the $p$-value of the K--S test, $p(\alpha
)$, since $p(\cdot)$ is a strictly decreasing function of $D(\cdot)$.

\begin{figure}[b]

\includegraphics{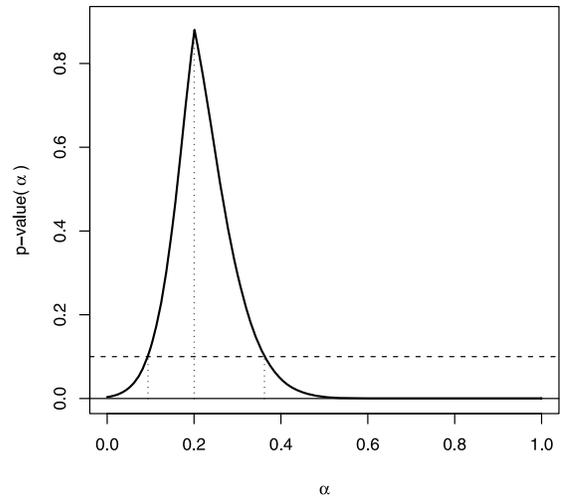}

\caption{Plot of the $p$-value of the K--S test as a function of the
parameter, showing the point estimate and the 90\% confidence interval,
for model A.}
\label{kspA}
\end{figure}

\begin{figure}

\includegraphics{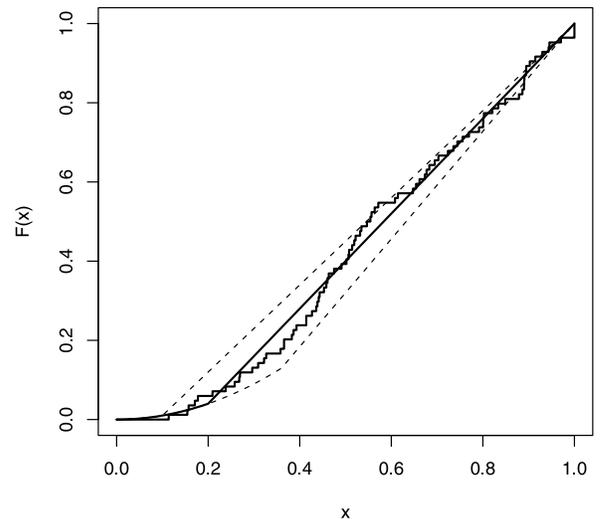}

\caption{Empirical cumulative distribution function of the $p$-values and
fitted model (solid line: $\hat{\alpha}=0.201$; dashed lines: 90\%
confidence limits).}
\label{pval3}
\end{figure}

Figure~\ref{kspA} shows the plot of the K--S $p$-values, $p(\alpha)$, as a
function of $\alpha$, together with the point estimate, $\hat{\alpha
}=0.201$ ($D=0.0623$, $P=0.8804$), and a 90\% confidence interval for
$\alpha,  (0.094;0.362)$. A detailed explanation on how these figures
were obtained is given in Appendix~\ref{apA3}. Figure~\ref{pval3}
confirms the good model fit.

This model can also be submitted to Fisher's chi-square analysis.
Assuming it holds for a certain value $\alpha_0$, we may still compute
``chi-square statistics,'' but the $p$-values can no longer be obtained
from the chi-square distribution. However, they can be accurately
estimated by Monte Carlo simulation. The difference to the previous
simulations is that statistics and ($\chi^2_1$) $p$-values were always
computed (for each of the 84 binomial cases and each random repetition)
and whenever that $p$-value was smaller than $\alpha_0$ another binomial
result was generated and the statistic recorded was the minimum of the two.

The simulation results obtained for three values of $\alpha$ (point
estimate and limits of the confidence interval) are presented in the
three columns of Table~\ref{Tab5} under the heading ``Model A.''
The $p$-values in these columns (and especially those corresponding to
$\alpha=0.201$) do not show any sign of being too close to one anymore,
in fact, they are perfectly reasonable. In Appendix~\ref{apB} we
present a more detailed (and technical) justification of the results obtained.

\begin{figure}

\includegraphics{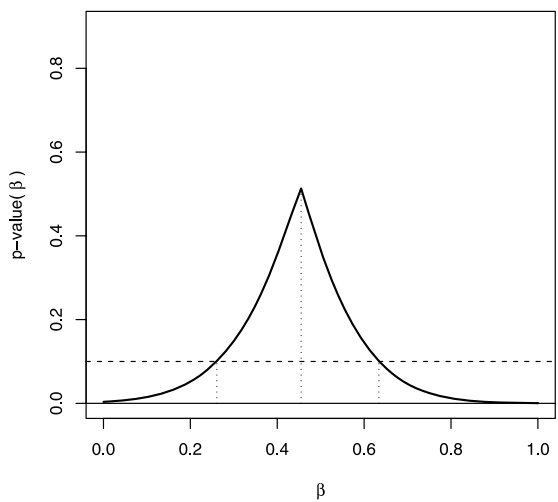}

\caption{Plot of the $p$-value of the K--S test as a function of the
parameter, showing the point estimate and the 90\% confidence interval,
for model B.}
\label{kspB}
\end{figure}

The conclusion is that our model explains Fisher's chi-square results:
Mendel's data are ``\textit{too good to be true}'' according to the
assumption that all the data presented in Mendel's paper correspond to
all the experiments Mendel performed, or to a random selection from all
the experiments. When this assumption is replaced by model A the
results can no longer be considered too good.
So we conclude that model A is a reasonable statistical explanation for
the controversy. We do not pretend that it is necessarily the ``true''
model; however, it is very simple and does provide extra insight into
the complexity of this historical debate and in this sense it is
useful. As G.E.P. Box said, ``\textit{All models are wrong, some models
are useful}.''

We have just seen how the suggested selection mechanism can make
Mendel's results (which we know are in fact correct) look too correct.
This raises a related question of general interest to all experimental
sciences: is it possible to make an incorrect theory look correct by
applying this or a similar selection mechanism? Although a detailed
answer to this question is beyond the scope of this paper, in
Appendix~\ref{apD} we give an idea on how a generalization of model A
can be used to explore the question.

\subsection{Alternative Models}\label{sec5.2}
No doubt there are many models, perhaps more complicated than ours,
that explain Mendel's data as well as, or perhaps better than, ours. A
relevant question to ask, then, is whether any model similar to ours,
more specifically, a one parameter model with c.d.f. varying between $x$
and $x^2$, would produce similar results and also a reasonable interpretation.

\begin{figure}[t]

\includegraphics{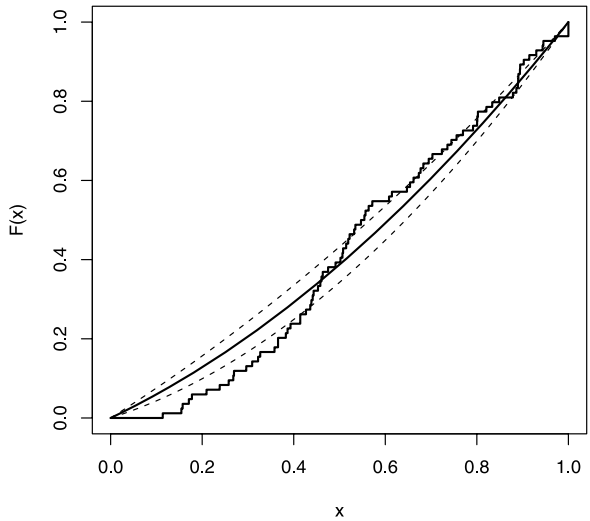}

\caption{E.c.d.f. of the $p$-values and alternative model (solid line: $\hat
{\beta}=0.45$; dashed lines: 90\% confidence limits).}
\label{pval4}
\end{figure}

To show that the answer to this question is negative, we have
considered an alternative model, model B, with distribution function
computed as a linear combination of the ``extreme'' models, that is,
with c.d.f. given by
$F_\beta(x)=(1-\beta)\times x + \beta\times x^2$, with $0 \le\beta
\le1$. This is mathematically simpler than model A and its
interpretation in terms of the design of the experiments could be:
Mendel would also decide to repeat some experiments and report only the
best result of both (the original and the repetition), but the decision
to repeat would be taken randomly with probability $\beta$, for
instance, by throwing a fair coin ($\beta=0.5$) or something similar.
Applying the methods described in Appendix~\ref{apA3} to this model,
we obtain (see Figure~\ref{kspB}) $\hat{\beta}=0.45$ (K--S test:
$D=0.0875$, $P=0.5131$) and $\mathit{CI}_{90\%}(\beta) = (0.261;0.634)$.
Figure~\ref{pval4} shows the e.c.d.f. of the $p$-values and the c.d.f. of model
B with $\beta=0.45$ (solid line) and $\beta= 0.261,  0.634$ (dashed
lines). Compared to Figure~\ref{pval3}, the fit of model B looks worse
than the fit of model A,
but it could still be considered acceptable. However, in what concerns
the chi-square analysis, model B is unable to produce good results
(cf. the last three columns in Table~\ref{Tab5}). The aggregated
$p$-value (84 $\mathit{df}$)
still points to ``\textit{too good to be true}'' except maybe for the
last column, which corresponds to the odd situation of randomly
repeating about 60\% of the experiments!
We have presented model B as just an exercise to show a specific
point, it does not correspond to a plausible procedure as does model A.

\begin{figure*}

\includegraphics{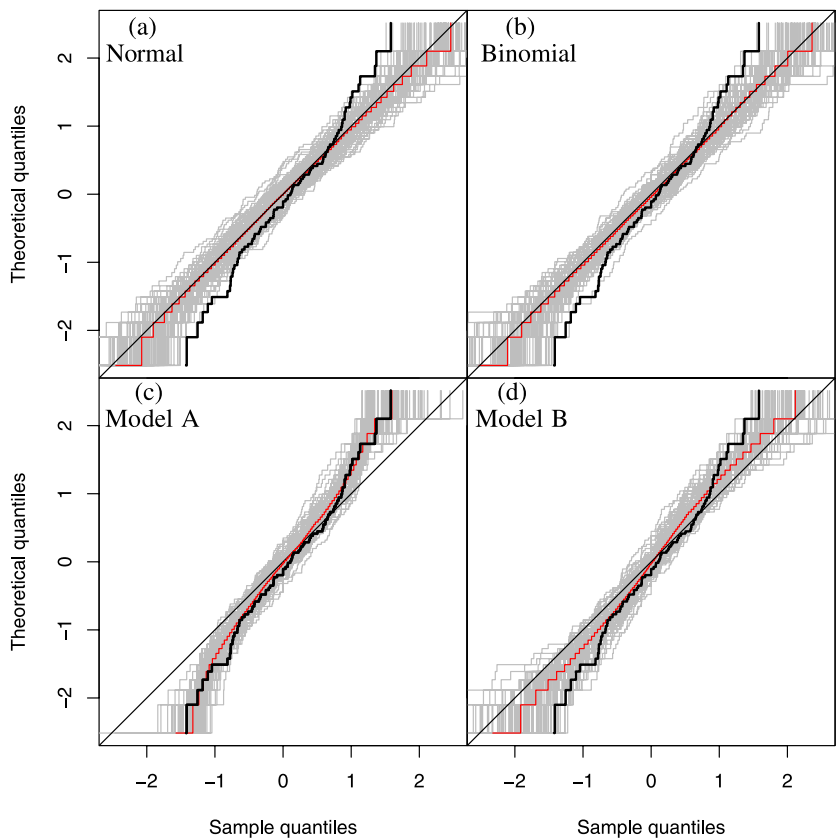}

\caption{Each part of the figure contains a normal quantile--quantile
plot of the original 84 Edwards' $\chi$ values (solid thick line), and
of 100 samples of simulated 84 $\chi$ values from each model (gray thin
lines) as well as an intermediate line, located in the ``middle'' of
the gray lines, corresponding to a ``synthetic'' sample obtained by
averaging the ordered observations of the 100 simulated samples.}
\label{Eplots}
\end{figure*}

\subsection{Further Support for the Proposed Model}\label{sec5.3}
As a harder challenge, we observed the behavior of each of the models
in the context of Edwards' chi values analysis, mentioned at the
beginning of Section~\ref{sec4.2}. The results of a simple simulation exercise
are represented in the four plots in Figure~\ref{Eplots}. All the plots
are normal quantile--quantile plots, and contain the representation of
the actual sample of 84 $\chi$ values (thick line). Each plot also
represents 100 samples of simulated 84 $\chi$ values, generated by the
corresponding model (gray thin lines), plus a ``synthetic'' sample
obtained by averaging the ordered observations of those 100 simulated
samples (intermediate line). Plot (a): the samples were generated from
a standard normal random variable (i.e., from the asymptotic
distribution of the $\chi$ values under the ideal assumptions, binomial
sampling and independent experiments). Plot (b): in this case the $\chi
$ values were obtained (by transformation of the $\chi^2$ values) from
the first 100 samples used to obtain the results given in the columns
with heading ``Edwards'' in Table~\ref{Tab5}. Plot (c): similar to the
previous but with the samples generated under model A. Plot~(d): idem
with model B.

From the top plots we conclude that the Normal and the Binomial models
are very similar and do not explain the observed values, whereas from
the bottom plots we can see that model A provides a much better
explanation of the $\chi$ values observed than model B. These
conclusions are no longer surprising, in the face of the previous
evidence; however, we shall remark that the several analyses are not
exactly equivalent, so the previous conclusions would not necessarily
imply this last one.

Besides the statistical evidence, which by itself may look speculative,
the proposed model is supported by Mendel's own words.
The following quotations from Mendel's paper (Mendel, \citeyear{Me1866}, page
numbers from Franklin et al., \citeyear{Fra2008}) are all relevant to our
interpretation:

\begin{quote}

``\textit{it appears to be necessary that all members of the series
developed in each successive generation should be, without exception,
subjected to observation}'' (page~80).

\end{quote}

From this sentence we conclude that Mendel was aware of the potential
bias due to incomplete observation, thus, it does not seem reasonable
that he would have deliberately censored the data or used sequential
sampling as suggested by some authors:

\begin{quote}

``\textit{As extremes in the distribution of the two seed characters
in one plant, there were observed in Expt. 1 an instance of 43 round
and only 2 angular, and another of 14 round and 15 angular seeds. In
Expt. 2 there was a case of 32 yellow and only 1 green seed, but also
one of 20 yellow and 19 green}'' (page~86).

\end{quote}

\begin{quote}

``\textit{Experiment 5, which shows the greatest departure, was
repeated, and then in lieu of the ratio of 60\,:\,40, that of 65\,:\,35
resulted}'' (page~89).

\end{quote}

Here he mentions repetition of an experiment but gives both results
(note that he decided to repeat an experiment with $p$-value${} =0.157$).
However, later he mentions several further experiments (pages~94, 95,
99, 100, 113) but presents results in only one case (page~99) and in
another suggests that the results were not good (page~95):

\begin{quote}

``\textit{In addition, further experiments were made with a smaller
number of experimental plants in which the remaining characters by twos
and threes were united as hybrids: all yielded approximately the same
results}'' (page~94).

\end{quote}

\begin{quote}

``\textit{An experiment with peduncles of different lengths gave on
the whole a fairly satisfactory results, although the differentiation
and serial arrangement of the forms could not be effected with that
certainty which is indispensable for correct experiment}'' (page~95).

\end{quote}

\begin{quote}

``\textit{In a further experiment the characters of flower-color and
length of stem were experimented upon}\ldots '' in this case results are
given, and then concludes ``\textit{The theory adduced is therefore
satisfactorily confirmed in this experiment also}'' (pages~99/100).

\end{quote}

\begin{quote}

``\textit{For the characters of form of pod, color of pod, and
position of flowers, experiments were also made on a small scale and
results obtained in perfect agreement}'' (page~100).

\end{quote}

\begin{quote}

``\textit{Experiments which in this connection were carried out with
two species of Pisum \ldots\  The two experimental plants differed in 5
characters,} \ldots '' (page~113).

\end{quote}

It is likely that the results omitted were worse and that Mendel may
have thought there would be no point in showing them anymore (he gave
examples of bad fit to his theory before, page~86).

Our model may be seen as an approximation for the omissions described
by Mendel. In conclusion, an unconscious bias may have been behind the
whole process of experimentation and if that is accepted, then it
explains the paradox and ends the controversy at last.

\section{Conclusion}\label{sec6}

Gregor Mendel is considered by many a creative genius and
incontestably the founder of genetics. However, as with many
revolutionary ideas, his laws of heredity (Mendel, \citeyear{Me1866}), a brilliant
and impressive achievement of the human mind, were not immediately
recognized, and stayed dormant for about 35 years, until they were
rediscovered in 1900. When Ronald Fisher, famous statistician and
geneticist, considered the father of modern statistics, used a
chi-square test to examine meticulously the data that Mendel had
provided in his classical paper to prove his theory, he concluded that
the data was too close to what Mendel was expecting, suggesting that
scientific misconduct had been present in one way or another. This
profound conflict raised a longstanding controversy that has been
engaging the scientific community for almost a century. Since none of
the proposed explanations of the conflict is satisfactory, a large
number of arguments, ideas and opinions of various nature (biological,
psychological, philosophical, historical, statistical and others) have
been continually put forward just like a never ending saga.

This study relies on the particular assumption that the experimentation
leading to the data analyzed by Fisher was carried out under a specific
unconscious bias. The argument of unconscious bias has been considered
a conceivable justification by various authors who have committed
themselves to study some variations of this line of reasoning
(Root-Bernstein, \citeyear{Ro1983};
Bowler, \citeyear{Bo};
Dobzhansky, \citeyear{Dob1967};
Olby, \citeyear{Ol1984};
Meijer, \citeyear{Mei1983};
Rosenthal, \citeyear{Ros1976};
Thagard, \citeyear{Th1988};
Nissani and
Hoefler-Nissani, \citeyear{Ni1992}). But all these attempts are based on somehow
subjective interpretations and throw no definite light on the problem.
On the contrary,
in this paper the type of unconscious bias is clearly identified
and a well-defined statistical analysis based on a proper statistical
model is performed. The results show that the model is a plausible
statistical explanation for the controversy.

The study goes
as follows: (i) Fisher results were confirmed by repeating his analysis
on the same real set of data and on simulated data, (ii) inspired by
Edwards' (\citeyear{Ed1986a}) approach, we next idealized a convenient model of a
sequence of binomial experiments and recognized that the $p$-value
produced by this model shows a slight increase, although it keeps very
close to the result obtained by Fisher. This gave us confidence to work
with this advantageous structure, (iii) we focused on the analysis of
the $p$-values of the previous model and realized that the $p$-values do
not have a uniform distribution as they should, (iv) the question arose
of what the distribution of the $p$-values could be, and we arrived at
the satisfactory model we propose in the text, (v) finally, assuming
that our model holds, and repeating the chi-square analysis adopted by
Fisher, one sees that the impressive
effect detected by Fisher disappears.

Returning to Fisher's reaction to the paradoxical situation he
encountered, one may think that, despite his remarkable investigation
(Fisher, \citeyear{Fi1936}) of Mendel's work, to prove that something had gone wrong
with the selection of the experimental data, apparently neither did he
question how could the data have been generated nor did he identify the
defects of the sample or give a statistical explanation for the awkward
result. In the end Fisher left an inescapable global impression of
scientific malpractice, a conclusion that he based on a sound
statistical analysis.

Probity is an essential component of the scientific work that should
always be contemplated to guarantee credible final results and
conclusions. That is why all measures should be taken to make sure that
neither conscious nor unconscious bias will affect the results of the
research work. Unfortunately there exists unconscious bias, an
intrinsic automatic human drive based on culture, social prejudice or
motivation that is difficult to stop. Hidden bias influences many
aspects of our decisions, our social behaviour and our work. That is
why scientific enterprises including honourable doctors and
well-intentioned patients do not dispense the scientific techniques
based on blind or double blind procedures. In Mendel's case we all know
that there was a profound motivation that could have triggered the bias
and in those days we guess that the attention given to unconscious bias
may have been poor or it may have not existed at all. Frequently
science ends up in detecting errors or fraud that have been induced by
bias. But there are no errors in Mendel's laws, or are there? So why
are we worried? Anyway, we wish that Mendel's unconscious bias
coincides with the arrangement we are suggesting in this paper, because
if Mendel did what we think he did, the controversy is finally over.

\begin{appendix}

\section{The 84 Binomial Experiments}\label{apC}

Edwards (\citeyear{Ed1986a}) organized Mendel's data as the result of 84 binomial
experiments. Note that this involves decomposition of the multinomial
experiments.
In this study we have relied on Edwards' decompositions. In order to
remain as close as possible to Fisher's choices, the data from
Table~\ref{Tab1}---already binomial experiments---were included exactly
as\break shown, unlike Edwards who subtracted the ``plant illustrations''
from these data.
According to this procedure, experiment No. 1 (No. 2) is not
independent of the ``plant illustrations'' Nos 8--17 (Nos 18--27). But, attending
to the relative magnitude of the number of
observations, if we had used Edwards' numbers, the final results would
have not been too different and the conclusions would have been the same.
We have also considered the theoretical ratio of $2:1$ throughout the
experiments involving $(F_3)$ generations, instead of the ratio
$0.63:0.37$ that Edwards used in some cases.
The number of binomial experiments per pair of true probabilities
(ratio) is as follows: 42 cases with $0.75:0.25$ ($3:1$); 15 cases with
$0.5:0.5$ ($1:1$); 27 cases with $2/3:1/3$ ($2:1$).

Table \ref{Tab84} contains the following information about the 84
binomial experiments considered in this paper:

\begin{description}
\item[Trait:] binary variable under consideration (the category of
interest is called a ``success'' and the other category is a
``failure'') using the following coding:
A (seed shape, round or wrinkled), B (seed color, yellow or green), C
(flower color, purple or white), D (pod shape, inflated or
constricted), E (pod color, yellow or green), F (flower position, axial
or terminal), G (stem length, long or short). The usual notation is
used to distinguish phenotype (italic inside quotation marks) from
genotype (italic), and the dominant form (upper case) from the
recessive (lowercase); see also Section~\ref{sec3} and Table \ref{Tab1}.

\item[$\bolds{n}$:] number of observations (Bernoulli trials) of
the experiment.

\item[Observed:] observed frequencies of ``successes'' ($n_1$) and
``failures'' ($n- n_1$). Under the standard assumptions $n_1 \sim
\operatorname{Bin}(n, p)$, where $p$ is the probability of a ``success'' in one
trial.

\item[$\bolds{p_0}$:] theoretical probability of a ``success''
under\break  Mendel's theory ($H_0\dvtx p=p_0$).

\item[$\bolds{\chi}$:] observed value of the test statistic to test $H_0$ against $H_1\dvtx p
\ne p_0$, given by $(n_1 -np_0)/\break\sqrt{np_0 (1 - p_0)}$.

\item[$\bolds p$-value:] $p$-value of the test. Assuming $n$ is large, $p$-value${}
= P  ( \chi_1^2 > \chi^2  )$.
\end{description}

\renewcommand{\thetable}{\arabic{table}}
\setcounter{table}{6}
\begin{table*}
\tabcolsep=0pt
\caption{Data from the 84 binomial experiments}\label{Tab84}
\begin{tabular*}{350pt}{@{\extracolsep{\fill}}ld{2.0}cd{4.0}d{4.0}d{4.0}cd{2.3}c@{}}
\hline
\multirow{2}{46pt}{\textbf{Type of experiment}} & & & & \multicolumn{2}{c}{\textbf{Observed}}  \\
 \cline{5-6}
 & \multicolumn{1}{c}{\textbf{No.}} & \multicolumn{1}{c}{\textbf{Trait}} &
 \multicolumn{1}{c}{$\bolds n$} & \multicolumn{1}{c}{$\bolds{n_{1}}$} &
\multicolumn{1}{c}{$\bolds{n-n_{1}}$} & \multicolumn{1}{c}{$\bolds{p_0}$} &
\multicolumn{1}{c}{$\bolds\chi$} &
\multicolumn{1}{c@{}}{$\bolds p$\textbf{-value}} \\
\hline
Single trait& 1 & A & 7324 & 5474 & 1850 & 3$/$4 & -0.513 & 0.608 \\
$F_2$& 2 & B & 8023 & 6022 & 2001 & 3$/$4 & 0.123 & 0.903 \\
& 3 & C & 929 & 705 & 224 & 3$/$4 & 0.625 & 0.532 \\
 & 4 & D & 1181 & 882 & 299 & 3$/$4 & -0.252 & 0.801 \\
& 5 & E & 580 & 428 & 152 & 3$/$4 & -0.671 & 0.502 \\
& 6 & F & 858 & 651 & 207 & 3$/$4 & 0.591 & 0.554 \\
& 7 & G & 1064 & 787 & 277 & 3$/$4 & -0.779 & 0.436 \\
[3pt]
Illustrations& 8 & A & 57 & 45 & 12 & 3$/$4 & 0.688 & 0.491 \\
of plant&9 & A & 35 & 27 & 8 & 3$/$4 & 0.293 & 0.770 \\
variation& 10 & A & 31 & 24 & 7 & 3$/$4 & 0.311 & 0.756 \\
$F_2$& 11 & A & 29 & 19 & 10 & 3$/$4 & -1.179 & 0.238 \\
& 12 & A & 43 & 32 & 11 & 3$/$4 & -0.088 & 0.930 \\
& 13 & A & 32 & 26 & 6 & 3$/$4 & 0.817 & 0.414 \\
& 14 & A & 112 & 88 & 24 & 3$/$4 & 0.873 & 0.383 \\
 & 15 & A & 32 & 22 & 10 & 3$/$4 & -0.817 & 0.414 \\
& 16 & A & 34 & 28 & 6 & 3$/$4 & 0.990 & 0.322 \\
& 17 & A & 32 & 25 & 7 & 3$/$4 & 0.408 & 0.683 \\
& 18 & B & 36 & 25 & 11 & 3$/$4 & -0.770 & 0.441 \\
& 19 & B & 39 & 32 & 7 & 3$/$4 & 1.017 & 0.309 \\
& 20 & B & 19 & 14 & 5 & 3$/$4 & -0.133 & 0.895 \\
& 21 & B & 97 & 70 & 27 & 3$/$4 & -0.645 & 0.519 \\
& 22 & B & 37 & 24 & 13 & 3$/$4 & -1.424 & 0.155 \\
& 23 & B & 26 & 20 & 6 & 3$/$4 & 0.227 & 0.821 \\
& 24 & B & 45 & 32 & 13 & 3$/$4 & -0.603 & 0.547 \\
& 25 & B & 53 & 44 & 9 & 3$/$4 & 1.348 & 0.178 \\
& 26 & B & 64 & 50 & 14 & 3$/$4 & 0.577 & 0.564 \\
& 27 & B & 62 & 44 & 18 & 3$/$4 & -0.733 & 0.463 \\
 [3pt]
Bifactorial&28 & A & 556 & 423 & 133 & 3$/$4 & 0.588 & 0.557 \\
experiment&29 & B among ``\textit{A}'' & 423 & 315 & 108 & 3$/$4 &
-0.253 & 0.801 \\
$F_2$&30 & B among ``\textit{a}'' & 133 & 101 & 32 & 3$/$4 & 0.250 &
0.802 \\
 [3pt]
Trifactorial&31 & A & 639 & 480 & 159 & 3$/$4 & 0.069 & 0.945 \\
experiment&32 & B among ``\textit{A}'' & 480 & 367 & 113 & 3$/$4 & 0.738 & 0.461 \\
$F_2$&33 & B among ``\textit{a}'' & 159 & 122 & 37 & 3$/$4 & 0.504 & 0.615 \\
&34 & C among \textit{AaBb} & 175 & 127 & 48 & 3$/$4 & -0.742 & 0.458 \\
&35 & C among \textit{AaBB} & 70 & 52 & 18 & 3$/$4 & -0.138
& 0.890 \\
&36 & C among \textit{AABb} & 78 & 60 & 18 & 3$/$4 & 0.392 &
0.695 \\
&37 & C among \textit{AABB} & 44 & 30 & 14 & 3$/$4 & -1.045 &
0.296 \\
&38 & C among \textit{Aabb} & 76 & 60 & 16 & 3$/$4 & 0.795 & 0.427 \\
&39 & C among \textit{AAbb} & 37 & 26 & 11 & 3$/$4 & -0.664 & 0.506 \\
&40 & C among \textit{aaBb} & 79 & 55 & 24 & 3$/$4 & -1.104 & 0.269 \\
&41 & C among \textit{aaBB} & 43 & 33 & 10 & 3$/$4 & 0.264 & 0.792 \\
&42 & C among \textit{aabb} & 37 & 30 & 7 & 3$/$4 & 0.854 & 0.393 \\
 [3pt]
Single trait&43 & A & 565 & 372 & 193 & 2$/$3 & -0.417 & 0.677 \\
($F_3$)&44 & B & 519 & 353 & 166 & 2$/$3 & 0.652 & 0.515 \\
&45 & C & 100 & 64 & 36 & 2$/$3 & -0.566 & 0.572 \\
&46 & D & 100 & 71 & 29 & 2$/$3 & 0.919 & 0.358 \\
&47 & E & 100 & 60 & 40 & 2$/$3 & -1.414 & 0.157 \\
&48 & F & 100 & 67 & 33 & 2$/$3 & 0.071 & 0.944 \\
&49 & G & 100 & 72 & 28 & 2$/$3 & 1.131 & 0.258 \\
&50 & E & 100 & 65 & 35 & 2$/$3 & -0.354 & 0.724 \\
\hline
\end{tabular*}
\end{table*}
\setcounter{table}{6}
\begin{table*}
\tabcolsep=0pt
\caption{(Continued)}
\begin{tabular*}{350pt}{@{\extracolsep{\fill}}ld{2.0}cd{3.0}d{3.0}d{3.0}cd{2.3}c@{}}
\hline
\multirow{2}{46pt}{\textbf{Type of experiment}} & & & & \multicolumn{2}{c}{\textbf{Observed}}  \\
 \cline{5-6}
 & \multicolumn{1}{c}{\textbf{No.}} & \multicolumn{1}{c}{\textbf{Trait}} &
 \multicolumn{1}{c}{$\bolds n$} & \multicolumn{1}{c}{$\bolds{n_{1}}$} &
\multicolumn{1}{c}{$\bolds{n-n_{1}}$} & \multicolumn{1}{c}{$\bolds{p_0}$} &
\multicolumn{1}{c}{$\bolds\chi$} &
\multicolumn{1}{c@{}}{$\bolds p$\textbf{-value}} \\
\hline
Bifactorial&51 & A among ``\textit{AB}'' & 301 & 198 & 103 & 2$/$3 & -0.326 &
0.744 \\
experiment&52 & A among ``\textit{Ab}'' & 102 & 67 & 35 & 2$/$3 &
-0.210 & 0.834 \\
($F_3$)&53 & B among ``\textit{aB}'' & 96 & 68 & 28 & 2$/$3 & 0.866 &
0.386 \\
&54 & B among \textit{Aa}``\textit{B}'' & 198 & 138 & 60 & 2$/$3 &
0.905 & 0.366 \\
&55 & B among \textit{AA}``\textit{B}'' & 103 & 65 & 38 & 2$/$3 &
-0.766 & 0.443 \\
 [3pt]
Trifactorial&56 & A among ``\textit{AB}'' & 367 & 245 & 122 & 2$/$3 & 0.037 & 0.971 \\
experiment&57 & A among ``\textit{Ab}'' & 113 & 76 & 37 & 2$/$3 & 0.133 & 0.894 \\
($F_3$)&58 & B among ``\textit{aB}'' & 122 & 79 & 43 & 2$/$3 & -0.448 & 0.654
\\
&59 & B among \textit{Aa}``\textit{B}'' & 245 & 175 & 70 & 2$/$3 & 1.581
& 0.114 \\
&60 & B among \textit{AA}``\textit{B}'' & 122 & 78 & 44 & 2$/$3 &
-0.640 & 0.522 \\
&61 & C among \textit{AaBb} & 127 & 78 & 49 & 2$/$3 &
-1.255 & 0.210 \\
&62 & C among \textit{AaBB} & 52 & 38 & 14 & 2$/$3 & 0.981 &
0.327 \\
&63 & C among \textit{AABb} & 60 & 45 & 15 & 2$/$3 & 1.369 & 0.171
\\
&64 & C among \textit{AABB} & 30 & 22 & 8 & 2$/$3 & 0.775 & 0.439 \\
&65 & C among \textit{Aabb} & 60 & 40 & 20 & 2$/$3 & 0.000 & 1.000 \\
&66 & C among \textit{AAbb} & 26 & 17 & 9 & 2$/$3 & -0.139 & 0.890 \\
&67 & C among \textit{aaBb} & 55 & 36 & 19 & 2$/$3 & -0.191 & 0.849 \\
&68 & C among \textit{aaBB} & 33 & 25 & 8 & 2$/$3 & 1.108 & 0.268 \\
&69 & C among \textit{aabb} & 30 & 20 & 10 & 2$/$3 & 0.000 & 1.000 \\
[3pt]
Gametic&70 & A & 90 & 43 & 47 & 1$/$2 & -0.422 & 0.673 \\
ratios&71 & B among \textit{AA} & 43 & 20 & 23 & 1$/$2 & -0.458 & 0.647 \\
&72 & B among \textit{Aa} & 47 & 25 & 22 & 1$/$2 & 0.438 & 0.662 \\
&73 & A & 110 & 57 & 53 & 1$/$2 & 0.381 & 0.703 \\
&74 & B among \textit{Aa} & 57 & 31 & 26 & 1$/$2 & 0.662 & 0.508 \\
&75 & B among \textit{aa} & 53 & 27 & 26 & 1$/$2 & 0.137 & 0.891 \\
&76 & A & 87 & 44 & 43 & 1$/$2 & 0.107 & 0.915 \\
&77 & B among \textit{AA} & 44 & 25 & 19 & 1$/$2 & 0.905 & 0.366 \\
&78 & B among \textit{Aa} & 43 & 22 & 21 & 1$/$2 & 0.153 & 0.879 \\
&79 & A & 98 & 49 & 49 & 1$/$2 & 0.000 & 1.000 \\
&80 & B among \textit{Aa} & 49 & 24 & 25 & 1$/$2 & -0.143 & 0.886 \\
&81 & B among \textit{aa} & 49 & 22 & 27 & 1$/$2 & -0.714 & 0.475 \\
&82 & G & 166 & 87 & 79 & 1$/$2 & 0.621 & 0.535 \\
&83 & C among \textit{Gg} & 87 & 47 & 40 & 1$/$2 & 0.751 & 0.453 \\
&84 & C among \textit{gg} & 79 & 38 & 41 & 1$/$2 & -0.338 & 0.736 \\
\hline
\end{tabular*}
\end{table*}
%

\section{Technical Details}

\subsection{Simulation of the Chi-Square Analysis} \label{apA1}
In each of the 1,000,000 repetitions a replicate of Mendel's complete
data set was generated, using the probabilities corresponding to the
theoretical ratios, and multinomial distributions with the appropriate
number of categories (which reduces to the binomial distribution for
the experiments with two categories and is strictly multinomial for the
remaining, bifactorial, trifactorial and gametic ratios). For each
replicate, a total ``chi-square'' statistic was computed as Fisher did
for the actual data set. From the 1,000,000 replicates of the test
statistic it is possible to estimate the $p$-value of the test without
knowledge of the sampling distribution of the test statistic.
Recall that the MC estimate of a $p$-value (or simulated $p$-value)
associated to a certain observed statistic (which increases as the data
deviate from the null hypothesis) is the number of repetitions for
which the corresponding simulated statistic is larger than the observed
statistic ($\chi^2_{\mathrm{obs}}$), divided by the number of repetitions. If we
denote an MC estimate of a $p$-value by $P$, the corresponding
estimated standard error is $\mathit{se}= \sqrt{P(1-P)/B}$, where $B$ is the
number of random repetitions. These figures are also reported in
Table~\ref{Tab5}.

\subsection{Analysis of the $p$-Values} \label{apA2}

The Kolmogorov--Smirnov (K--S) test is a goodness-of-fit test based on
the statistic $D=\sup_x  | F_n(x) - F_0 (x)  |$,
where $F_n(x)$ is the e.c.d.f. obtained from a random sample $(x_1, \ldots
, x_n)$ and $F_0 (x)$ is a hypothesized, completely specified, c.d.f. [$D$
is simply the largest vertical distance between the plots of $F_n(x)$
and $F_0 (x)$]. This test was selected for analyzing the c.d.f. of the
$p$-values because it is more powerful for detecting deviations from a
continuous distribution than other alternatives such as the chi-square
goodness-of-fit test (Massey, \citeyear{Ma1951}). Under the appropriate conditions
[$F_0 (x)$ is continuous, there are no ties in the sample], the exact
$p$-value of the K--S test can be computed. In our analysis these
conditions are not exactly met (the true c.d.f. is not continuous and
because of that there are ties in the data), so it is necessary to
proceed with caution.

The first K--S test performed intended to test the uniformity of the 84
$p$-values and produced $D=0.1913$ ($P=0.0036$). The ``exact'' $p$-value
was computed after eliminating the ties by addition of a small amount
of noise to each data point (random numbers generated from a normal distribution
with zero mean and standard deviation $10^{-7}$).

As there are several approximations involved, we checked the whole
procedure by performing a simulation study similar to the one described
in Section~\ref{sec4.1} for the chi-square analysis. In 1,000,000 random
repetitions of the sample of 84 $p$-values
a simulated $p$-value of 0.0038 ($\mathit{se}=0.00006$) was obtained (the K--S
statistic was larger than 0.1903 in 3807 repetitions). This is
statistically significantly larger than 0.0036; however, the difference
is not meaningful from a practical point of view, the ``exact'' $p$-value
is 3 digits accurate. So we concluded that it is acceptable to use the
K--S test as described.

\setcounter{table}{7}
\begin{table*}
\tabcolsep=0pt
\caption{Illustration of the computations necessary to obtain the exact distribution
of the $p$-values ($n=35, p=0.75$)}\label{tab8}
\begin{tabular*}{\textwidth}{@{\extracolsep{\fill}}ld{3.2}d{3.2}cd{2.3}d{2.4}
d{2.3}cd{2.3}d{2.4}d{2.4}@{}}
  \hline
$y$ &  0 & 1 &\ldots & 25 & 26 & 27 & \ldots & 33 & 34 & 35 \\
$\chi^2(y)$ &  105.00 & 97.15 &
\ldots & 0.24 & 0.0095 & 0.086 & \ldots & 6.94 & 9.15 & 11.67 \\
 $p$-value  &  \multicolumn{1}{c}{\hphantom{10}10$^{-24}$} &
 \multicolumn{1}{c}{\hphantom{10}10$^{-22}$} &
\ldots &
0.626 & 0.922 & 0.770 & \ldots &
0.008 & 0.002 & 0.0006 \\
$P(y)$ &  \multicolumn{1}{c}{\hphantom{10}10$^{-21}$} &
\multicolumn{1}{c}{\hphantom{10}10$^{-19}$} &
\ldots & 0.132 & 0.152 & 0.152 & \ldots &
0.003 & 0.0005 & \multicolumn{1}{c}{$10^{-5}$\hphantom{,0}}\\
\hline
\end{tabular*}\vspace*{-2.5pt}
\end{table*}
\begin{table*}[b]
\tabcolsep=0pt
\caption{The exact distribution of the $p$-values, when  $n=35$ and $p=0.75$}\label{tab9}
\begin{tabular*}{\textwidth}{@{\extracolsep{\fill}}lccccccccc@{}}
  \hline
 $p$-value  & 0.001 & 0.002 & 0.005 & 0.008 & 0.015 & 0.025 & 0.040
& 0.064 & 0.097\\
 c.d.f.  & 0.002 & 0.003 & 0.007 & 0.010 & 0.019 & 0.029 & 0.050 &
0.077 & 0.117 \\
 $p$-value  & 0.143 & 0.205 & 0.283 & 0.380 & 0.495 & 0.626 & 0.770
& 0.922 \\
 c.d.f.  & 0.173 & 0.240 & 0.334 & 0.434 & 0.564 & 0.696 & 0.848 &
 1.000\\
 \hline
\end{tabular*}\vspace*{-2.5pt}
\end{table*}

There is another aspect which needs to be analyzed.
Because the outcomes of the experiments are binomial, yielding whole
numbers, the actual distribution of the $p$-values is discrete, not
uniform continuous. Therefore, we decided to investigate the
differences between the true distribution and the uniform continuous.
The exact distribution of the $p$-values obtained when the 84 chi-square
tests are applied to the binomial observations was determined in the
following way.

For a fixed experiment (with number of trials, $n$, and probability,
$p$) we can list the $n+1$ possible $p$-values along with the
corresponding probabilities. For instance, in one of the experiments
the number of seeds (trials) is $n= 35$ and the true probability of a
round seed is 0.75 (under Mendel's theory, i.e., the null hypothesis).
The possible values of round seeds observed in a repetition of this
experiment are 0, 1, 2, \ldots , 33, 34, 35 ($y$), each producing a
possible value of the chi-square statistic ($\chi^2(y)=(y-35\times
0.75)^2/(0.75 \times0.25 \times35)$) and a corresponding $p$-value $=
P (\chi^2_1 > \chi^2 (y) )$ with probability given by
$P(y)=C^{35}_y\times0.75^y \times0.25^{35-y}$ (see Table~\ref{tab8}).

Ordering the $p$-values and summing up the probabilities leads to the
discrete c.d.f. defined by the   points in Table~\ref{tab9}.

\renewcommand{\thefigure}{\arabic{figure}}
\setcounter{figure}{12}
\begin{figure}

\includegraphics{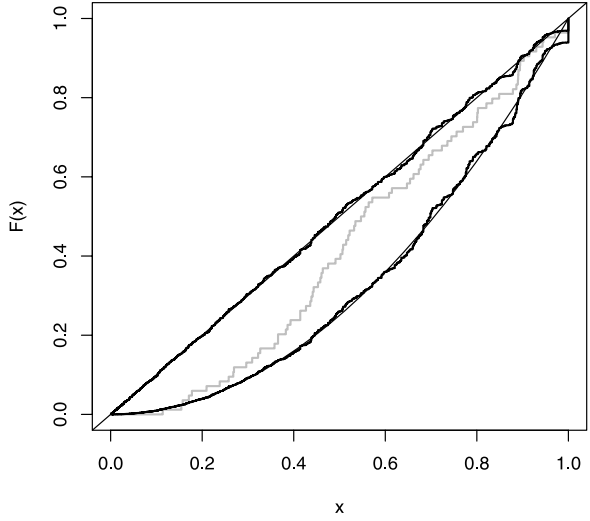}

\caption{Equivalent to Figure~\protect\ref{pval1} but showing the actual c.d.f. of
the $p$-values under binomial sampling (black stair steps line close to
the diagonal) and the c.d.f. of the maximum of two $p$-values (lower black
stair steps line).}
\label{pval2}
\end{figure}

 Proceeding similarly for all the 84 experiments and combining
the lists $P(y)$ multiplied by 1$/$84 (i.e., the contribution of each
experiment to the overall distribution), we obtain the global
probability function of the $p$-values, from which the final\vadjust{\goodbreak} cumulative
distribution is computed (overall there are 14,218 distinct possible
$p$-values, but from those the smallest 12,110 were not considered
because their cumulative probability is smaller than 0.001). The result
is shown in Figure~\ref{pval2}. Although for some of the experiments,
when considered individually, the c.d.f. of the $p$-values is quite
different from that of the uniform $(0,1)$ distribution (like in the
example above), when the 84 experiments are taken together the
resulting c.d.f. of the $p$-values is very close to the straight line
$F(x)=x$, which means that we can safely approximate this distribution
by a continuous uniform distribution in $(0,1)$ and trust the results
obtained with the K--S test (the approximation is not so good near the
upper right corner, but this area is not relevant to this conclusion).
The same remarks apply when the exact distribution of the maximum of
two $p$-values
is approximated by the curve $y=x^2$ (see also Figure~\ref{pval1}).

\subsection{Estimation of the Parameter of Model A} \label{apA3}

As explained in Section~\ref{sec5.1}, we consider the estimate of $\alpha$
defined as the value of $\alpha$ which maximizes the $p$-value of the K--S
test for testing the uniformity of the experimental $p$-values, denoted
by $p(\alpha)$.
The solution can be found by grid search, varying $\alpha$ in a finite
set of equidistant points between 0 and 1. With a grid width of 0.001,
the value $\hat{\alpha}=0.201$ was obtained. It is also possible
(Easterling, \citeyear{Eas1976}) to compute a $100 \times\gamma\%$ confidence
interval for $\alpha$ by inversion of the K--S test. This confidence
interval is the set of points $\alpha\in(0,1)$ such that $\mbox
{$p$-value}(\alpha)\ge1-\gamma$ (it may happen that this confidence set
is empty, which is an indication that the model is not appropriate).

A simulation study was performed to validate this procedure. 1000
samples of 84 $p$-values were generated from the 84 binomial experiments,
but considering the repetition mechanism of model A with $\alpha=0.2$.
For each of those 1000 samples the point estimate and the 90\%
confidence interval for $\alpha$ were computed as described in the
previous paragraph. The results of the simulation confirmed that the
whole procedure is adequate and performs as expected: the 1000 point
estimates are distributed almost symetrically with mean${}={}$0.2077
($\mathit{se}=0.0032$), median${}={}$0.194 and standard deviation${}={}$0.101. The
confidence set was empty in one case only. From the remaining 999
intervals (mean length${}={}$0.3019, $\mathit{se} =0.0049$; median length${}={}$0.272),
895 contained the true value of $\alpha=0.2$, which gives an estimated
confidence level of 89.5\%, in close agreement with the specified 90\%
confidence.

\section{The Chi-Square Analysis Assuming Model A} \label{apB}

The aim of this note is to show in detail why model A explains the
chi-square analysis, and to derive theoretically the approximate
distribution of the global chi-square statistic which can be used to
compute approximate $p$-values without the need to run simulations.

Fisher's chi-square analysis is based on the following simple reasoning:
Let $X_i$, $i=1, \ldots , 84$, be the random variable describing the
results of the $i$th experiment, that is, the number of observations
among $n_i$ which are classified into a category of interest (which one
of the two categories is the category of interest is not relevant). Let
$p_i$ be the probability of an observation of that category in a single
trial and $p_{i0}$ the value of the same probability according to
Mendel's theory. The standard model is $X_i \sim \operatorname{Bin}(n_i, p_{i})$. If,
furthermore, it is assumed, as Fisher did, that the $X_i$ are
independent and $H_{0i}\dvtx p_i=p_{i0}$ is true for all $i=1, \ldots ,
84$, it follows that
\begin{eqnarray*}
&&X_1, \ldots , X_{84} \stackrel{\mathrm{i.n.d.}}{\sim} \operatorname{Bin}(n_i, p_{i0})
\\
&& \quad \Rightarrow \quad
\chi_i=\frac{X_i- n_i p_{i0}}{\sqrt{n_i p_{i0} (1 - p_{i0})}}
\stackrel{\mathrm{i.i.d.}}{\sim}_{\!a}  N(0,1)   \\
 && \quad \Rightarrow \quad Q_i=\frac{(X_i- n_i p_{i0})^2}{n_i p_{i0} (1 - p_{i0})}  \stackrel
{\mathrm{i.i.d.}}{\sim}_{\!a}  \chi_1^2 \\
&& \quad  \Rightarrow  \quad    Q_T= \sum
_{i=1}^{84} Q_i   \sim_a   \chi_{84}^2.
\end{eqnarray*}
We also have that $E(Q_T)=84$ and $\operatorname{var} (Q_T)=168$ [$E(Q_i)=1$
and $\operatorname{var} (Q_i)=2$], and $p$-value $= P(Q_T>Q_{T\mathrm
{\ observed}})=P(Q_T> 41.3376)\simeq0.99998$.

From Mendel's paper we already know that he performed other experiments
than the 84 binomial experiments we have been considering.
Let us assume that he has (or could have) done $2 \times84 = 168$
binomial experiments, such that for each of the reported 84 experiments
there is a repetition (either actual or conceptual) and denote the
repetition of $X_i$ by $X_{i+84}$ and the corresponding chi-square
statistics by $Q_i$ and $Q_{i+84}$. If for each pair ($X_i, X_{84+i}$)
the selection of the reported experiment is random, then the observed
statistics, denoted $Q_i^\star$, $i=1, \ldots , 84$, are still i.i.d. $\chi
^2_1$ and Fisher's analysis remains valid. However, if the selection is
not random,
and is done according to our model A, we still have that (assuming, as
Fisher did, that $X_i$ are independent and $H_{0i}$: $p_i=p_{i0}$ is
true for all $i=1, \ldots , 168$)
\begin{eqnarray*}
&&X_1, \ldots , X_{168} \stackrel{\mathrm{i.n.d.}}{\sim} \operatorname{Bin}(n_i, p_{i0})
\\
&& \quad \Rightarrow \quad
\chi_1, \ldots , \chi_{168} \stackrel{\mathrm{i.i.d.}}{\sim}_{\!a}
N(0,1)\\
&& \quad \Rightarrow \quad
Q_1, \ldots , Q_{168} \stackrel{\mathrm{i.i.d.}}{\sim}_{\!a} \chi_1^2,
\end{eqnarray*}
but each of the observed statistics, $Q_i^\star$, $i=1, \ldots , 84$,
is no longer randomly chosen between $Q_i$ and $Q_{i+84}$, in fact,
they are chosen by the following rule,
\[
Q_i^\star=
\cases{\displaystyle Q_i  ,&   if   $Q_i \le c_\alpha$,\cr\displaystyle
\min(Q_i
, Q_{i+84}) ,&   if   $Q_i > c_\alpha$,
}
\]
where $c_\alpha$ is the $1-\alpha$ quantile of the $\chi^2_1$
distribution. Therefore, the $Q_i^\star$ are i.i.d. but do not follow the
$\chi^2_1$ distribution, and, in consequence,
$Q_T^\star= \sum_{i=1}^{84} Q_i^\star$ also does not follow the $\chi
^2_{84}$ distribution.

The exact distribution of $Q_T^\star$ appears to be very difficult to
derive; however, by the Central Limit Theorem (CLT), we can use a
normal approximation,
%
\begin{eqnarray}
\label{Qnormal}
Q_T^\star\sim_a N(84  \mu^\star, 84  \sigma^{\star2}),\nonumber
\\[-8pt]
\\[-8pt]
\eqntext{\mbox{with } \mu^\star=E(Q_i^\star) \mbox{ and } \sigma^{\star2} = \var
(Q_i^\star) .}
\end{eqnarray}
Assuming that $Q_i \sim\chi^2_1$, it is possible to compute the mean
and the variance of $Q_i^\star$, either directly or determining first
the pdf of $Q_i^\star$, $f_{Q^\star}$.

\setcounter{figure}{13}
\begin{figure*}

\includegraphics{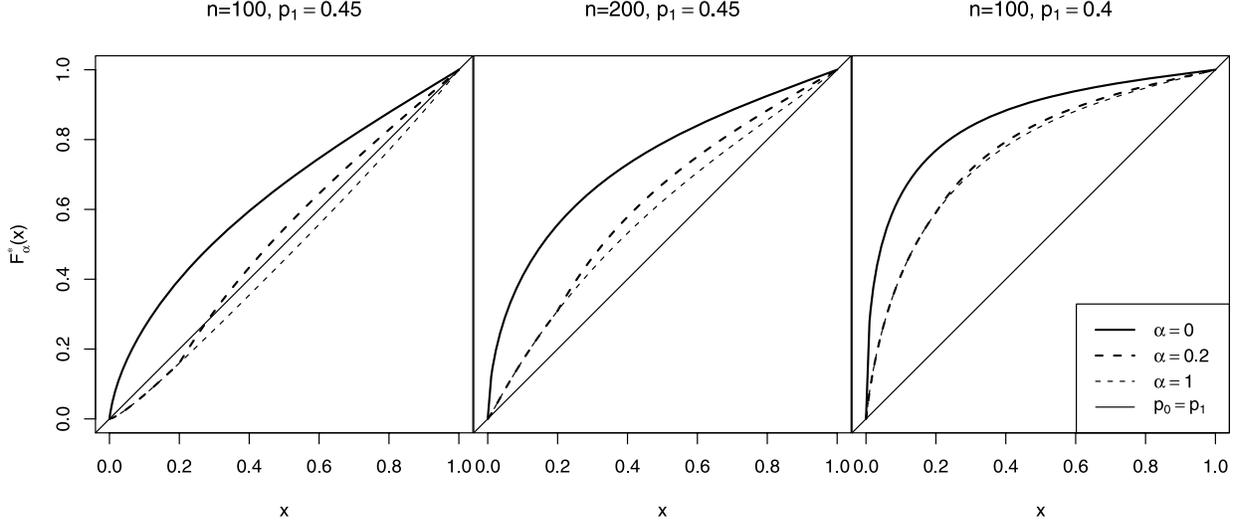}

\caption{Plots of $F^\ast_\alpha$ when $F_0$ is given by (\textup{\protect\ref{F0}}) for
$\alpha=0,0.2,1$, $p_0=0.5$ and three combinations of $(n,p_1)$.}
\label{figApD}
\end{figure*}

Given the reported value of the statistic, $Q^\star_i$ according to
model A, and the $p$-value computed using the chi-square distribution,
given by
$P=1-F_{Q_i}(Q^\star_i)$, with distribution function given by (\ref
{FmodelA}), we have that
\begin{eqnarray*}
F_{Q^\star}(x) &=& P(Q^\star\le x) = P \bigl( P \ge1- F_{Q_i}(x) \bigr)\\
&=& 1-F_P  \bigl( 1 - F_{Q_i}(x)  \bigr).
\end{eqnarray*}
Taking derivatives on both sides yields
\begin{eqnarray*}
f_{Q^\star}(x) &=& f_{Q_i}(x)   \frac{dF_P(u)}{du} \bigg |_{u=1-
F_{Q_i}(x)}\\
 &=&
\cases{\displaystyle  2 f_{Q_i} (x)  [ 1- F_{Q_i} (x)  ] ,&
 if   $x > c_\alpha$,\cr\displaystyle
 (1+ \alpha)f_{Q_i} (x)  ,&   if   $x \le
c_\alpha $,
}
\end{eqnarray*}
where $f_{Q_i} (x) = e^{-x/2}/\sqrt{2 \pi x}$, $x>0$, and $F_{Q_i} (x)
= \int_0^{x} f_{Q_i} (u)\,du$.

Using symbolic computation, we obtained
$\mu^\star=
1-  ( 2k_\alpha+ (1- \alpha) \sqrt{2c_\alpha k_\alpha}  )$,
$\sigma^{\star2} =
2-  ( 4k_\alpha^2 + (1-\alpha)\sqrt{2c_\alpha k_\alpha}  ( 4
k_\alpha+ 1 + c_\alpha ) + 2(2+c_\alpha(2 -2\alpha+\alpha
^2))k_\alpha )$,
with $k_\alpha= e^{-c_\alpha}/\pi$. Table~\ref{Tabapp} gives the
values of $\mu^\star$ and $\sigma^{\star2}$, as well as the $p$-values
obtained using the normal approximation (\ref{Qnormal}), for the three
values of $\alpha$ considered previously. The $p$-values obtained in the
simulation study (see Table~\ref{Tab5}) are also provided for
comparison. The two columns of $p$-values are very similar.
The results presented in this appendix are thus an independent
validation of the simulation results, in case there was any doubt about them.

\setcounter{table}{9}
\begin{table}
\tabcolsep=0pt
\caption{Mean and variance of $Q_i^\star$ and $p$-values obtained using
the normal approximation to $Q_T^\ast$ and from the Monte Carlo simulation}
\label{Tabapp}
\begin{tabular*}{\columnwidth}{@{\extracolsep{\fill}}lccccc@{}}
\hline
 &&&&$\bolds p$\textbf{-value}  &   $\bolds p$\textbf{-value}\\
$\bolds\alpha$ & $\bolds{c_\alpha}$ & $\bolds{\mu^\ast}$ & $\bolds{\sigma^{\star2}}$ &
\textbf{(normal approx.)} & \textbf{(simulation)}\\
\hline
0.094 & 2.805 & 0.6636 & 0.5685 & 0.9814 & 0.9860 \\
0.201 & 1.635 & 0.5160 & 0.3662 & 0.6412 & 0.6577 \\
0.362 & 0.831 & 0.4164 & 0.3135 & 0.1076 & 0.1176 \\
\hline
\end{tabular*}
\end{table}

\section{Model A for an Incorrect~Theory} \label{apD}

Suppose that Mendel's theory was not right but that the same selection
mechanism was applied (i.e., an experiment was repeated
whenever its $p$-value was smaller than $\alpha$, $0 \le\alpha\le1$,
and then only the experiment with the largest $p$-value was reported).
The difference between this case and that one considered in Section~\ref{sec5.1}
is that the original distribution of the $p$-values
is not uniform $(0,1)$ but has a c.d.f. $F_0(x)\ne x$ for some $0<x<1$.
Then, proceeding as in the proof of (\ref{FmodelA}),
we can conclude that the $p$-values effectively reported have a c.d.f.
given by
\[
F^\ast_\alpha(x) =
\cases{\displaystyle
 [F_0(x) ]^2, \cr\quad  \mbox{if }   0 \le x \le\alpha  ,\cr\displaystyle
 [1 + F_0( \alpha) ] F_0(x) - F_0(\alpha),  \cr\quad  \mbox{if }   \alpha
< x \le1.
}
\]
The selection procedure would make an incorrect theory look correct if
$F^\ast_\alpha(x)$ is ``close'' to the c.d.f. of a uniform $(0,1)$ random
variable. The result depends on the starting point, $F_0(x)$, which in
turn depends on the particular test under analysis and on the true and
hypothesized parameters, as the following example shows.

Suppose that the theory states that the success probability of a
binomial random variable is $p_0$ but that data are actually observed
from a binomial random variable with success probability $p_1$ which
may be different from $p_0$.
Assuming that $n$ is large, the normal approximation to the binomial
leads to
%
\begin{equation}
F_0(x) = \Phi \biggl( \frac{-z-\delta}{\eta}  \biggr) + 1 - \Phi \biggl(
\frac{z-\delta}{\eta}  \biggr),
\label{F0}
\end{equation}
where $\Phi(x)$ is the c.d.f. of a standard normal random variable, $z =
\Phi^{-1}  ( 1 - x/2  )$,
\[
\delta= \frac{n(p_1-p_0)}{\sqrt{np_0(1-p_0)}}   \quad \mbox{and} \quad
\eta^2 = \frac{p_1(1-p_1)}{p_0(1-p_0)}.
\]
Note that, when $p_0=p_1$, $F_0(x) \equiv x$, as it should.

Figure~\ref{figApD} shows the results for $p_0=1/2$ and some values of
$n$, $p_1$ and $\alpha$. We conclude that in the first case ($n=100$,
$p_1=0.45$) it is easy to make the theory look correct, but as $n$
increases or $p_1$ deviates from $p_0$
that becomes more difficult.

There is, of course, the possibility of further generalizing model A by
making more than 2 repetitions per experiment, say, $k$. With this
extra flexibility it is easy to make any theory look correct.

\end{appendix}

\section*{Acknowledgments}

The authors would like to thank the Editor, the Associate Editor and
two referees whose critical but constructive remarks and useful
suggestions have greatly improved the contents of this paper.


\begin{thebibliography}{99}
\bibitem[\protect\citeauthoryear{Pearson}{1900}]{Pe1900} \textsc
{Pearson, K.} (1900). On the criterion that a given system of
deviations from the probable in the case of a correlated system of
variables is such that it can be reasonably supposed to have arised
from random sampling. \textit{Philos. Mag.} \textbf{50} 157--175.\
\end{thebibliography}
\end{document}